\documentclass[a4paper,12pt]{article}
\pdfoutput=1

\usepackage[utf8]{inputenc}
\usepackage[T1]{fontenc}
\usepackage{dblfloatfix} 
\usepackage{fixltx2e} 
\usepackage{float}
\usepackage{wrapfig}
\usepackage{rotating}
\usepackage[normalem]{ulem}
\usepackage[font=small,labelfont=bf,labelsep=endash,margin=20pt]{caption}
\usepackage[title,header]{appendix}

\usepackage{relsize} 
\usepackage{amsmath}
\usepackage{marvosym}
\usepackage{wasysym}
\usepackage{amssymb}
\usepackage{latexsym}
\usepackage{physics}

\usepackage{bm}
\usepackage{mathtools}
\usepackage{newtxtext}\usepackage[utopia,greekuppercase=italicized]{mathdesign}\edef\partial{\mathchar\number\partial\noexpand\!}
\usepackage{isomath} 
\usepackage[margin=25mm]{geometry}
\usepackage{grffile}
\usepackage{enumitem} 
\setlist{itemsep=2pt,parsep=2pt,topsep=0pt,partopsep=0pt}
\usepackage{graphicx}
\usepackage{url}
\usepackage[hyperref,x11names]{xcolor}
\usepackage[colorlinks=true,linkcolor=Firebrick4,citecolor=blue,urlcolor=SteelBlue4]{hyperref}
\usepackage[authoryear]{natbib}
\bibpunct{(}{)}{;}{a}{}{,}
\usepackage{ctable} 
\usepackage{tabularx} 
\usepackage{mathbbol} 
\usepackage{bbm} 
\usepackage{thmtools} 
\usepackage{siunitx}

\def\cbeg{}
\def\cend{}

\renewcommand{\d}{\mathrm{d}}
\newcommand{\der}{\mathrm{d}}
\newcommand{\p}{\partial}
\newcommand{\pd}[2]{\frac{\partial #1}{\partial #2}}

\newcommand{\td}[2]{\frac{\der #1}{\der #2}}

\newcommand{\Order}{\mathcal{O}}
\renewcommand{\order}{\mathrm{o}}
\newcommand{\e}{\mathrm{e}}

\renewcommand{\b}[1]{{\bm{#1}}} 

\newcommand{\pvect}[2][r]{\begin{pmatrix*}[#1]#2\end{pmatrix*}} 

\renewcommand{\pmat}{\pvect}

\renewcommand{\geq}{\geqslant}\renewcommand{\leq}{\leqslant}\renewcommand{\ge}{\geqslant}
\renewcommand{\i}{\ensuremath{\text{i}}}

\newcommand{\keywords}[1]{\vspace{2mm}\noindent\textbf{Key words:} #1} 

\let\paragraphold\paragraph
\renewcommand*{\paragraph}[1]{\paragraphold{#1.}} 

\begin{document}
\title{\bf Mechanical cell interactions on curved interfaces}
\renewcommand{\thefootnote}{\fnsymbol{footnote}}%
\author{Pascal R Buenzli$^\text{a,}$\footnotemark[1]\ , Shahak Kuba$^\text{a}$, Ryan J Murphy$^\text{b}$, Matthew J Simpson$^\text{a}$}

\date{\small \vspace{-2mm}$^\text{a}$School of Mathematical Sciences, Queensland University of Technology (QUT), Brisbane, Australia\\$^\text{b}$School of Mathematics and Statistics, The University of Melbourne, Parkville, Australia\\\vskip 1mm \normalsize \today\vspace*{-5mm}}

\maketitle
\begin{abstract}
We propose a simple mathematical model to describe the mechanical relaxation of cells within a curved epithelial tissue layer represented by an arbitrary curve in two-dimensional space. \cbeg This model generalises previous one-dimensional models of flat epithelia to investigate the influence of curvature for mechanical relaxation. We represent the mechanics of a cell body either by straight springs, or by curved springs that follow the curve's shape. To understand the collective dynamics of the cells, we devise an appropriate continuum limit in which the number of cells and the length of the substrate are constant but the number of springs tends to infinity. In this limit, cell density is governed by a diffusion equation in arc length coordinates, where diffusion may be linear or nonlinear depending on the choice of the spring restoring force law. Our results have important implications about modelling cells on curved geometries: (i)~curved and straight springs can lead to different dynamics when there is a finite number of springs, but they both converge quadratically to the dynamics governed by the diffusion equation; (ii)~in the continuum limit, the curvature of the tissue does not affect the mechanical relaxation of cells within the layer nor their tangential stress; (iii)~a cell's normal stress depends on curvature due to surface tension induced by the tangential forces. Normal stress enables cells to sense substrate curvature at length scales much larger than their cell body, and could induce curvature dependences in experiments.\cend
	
\end{abstract}

\protect\footnotetext[1]{Corresponding author. Email address: \texttt{pascal.buenzli@qut.edu.au}}%
\renewcommand{\thefootnote}{\arabic{footnote}}%

\section{Introduction}
Epithelial tissues are composed of confluent arrangements of cells that interact mechanically with each other through contact-induced cell body deformations~\citep{ladoux2017}. Linking the mechanical properties of a tissue to those of its cells can help us understand how tissues grow and take their shape under spatial constraints as they interact with other tissues. It can also help us infer cell properties from experimentally observed tissue behaviour. Cells respond to mechanical cues, which may influence their proliferative behaviour, differentiation and survival~\citep{opas1989,weinans1996,nelson2005,keefer2011,moeendarbary2014,xi2019,nelson2022}. The development of detailed mathematical models of cell and tissue mechanics is important to understand how cells behave in response to mechanics in morphogenesis and developmental biology, tumour growth, wound healing, and tissue engineering~\citep{odell1981,roose2003,poujade2007,basan2010,murray2010,tse2012,idema2014,huang2014,cox2018,xi2019,flegg2020,murphy2021,karakaya2022,jensen2023}.

Tissue mechanics is intricately related to tissue geometry. The geometric arrangement of cells within a tissue affects their mechanical interactions. Numerous studies have shown that geometry has a strong influence on tissue growth rate both in-vivo and in tissue engineering constructs~\citep{rumpler2008,bidan2012a,bidan2012b,bidan2016,alias2017,alias2018,callens2020,callens2023,fratzl2022,schamberger2023}. This geometric control of tissue growth is in part attributed to changes in cell mechanical stress which influence cell behaviours through mechanobiological processes~\citep{nelson2005,luciano2021,zmurchok2018,tambyah2020,murphy2021}. Generally, it can be expected that where and how fast biological tissues grow and develop, is a combination of spatial constraints due to limited space availability, mechanical interactions, as well as biological cell behaviour~\citep{odell1981,dunlop2010,gamsjaeger2013,goriely2017,wang2018,ambrosi2019,cox2024}.

\begin{figure}[t!] \centering\includegraphics[scale=0.76]{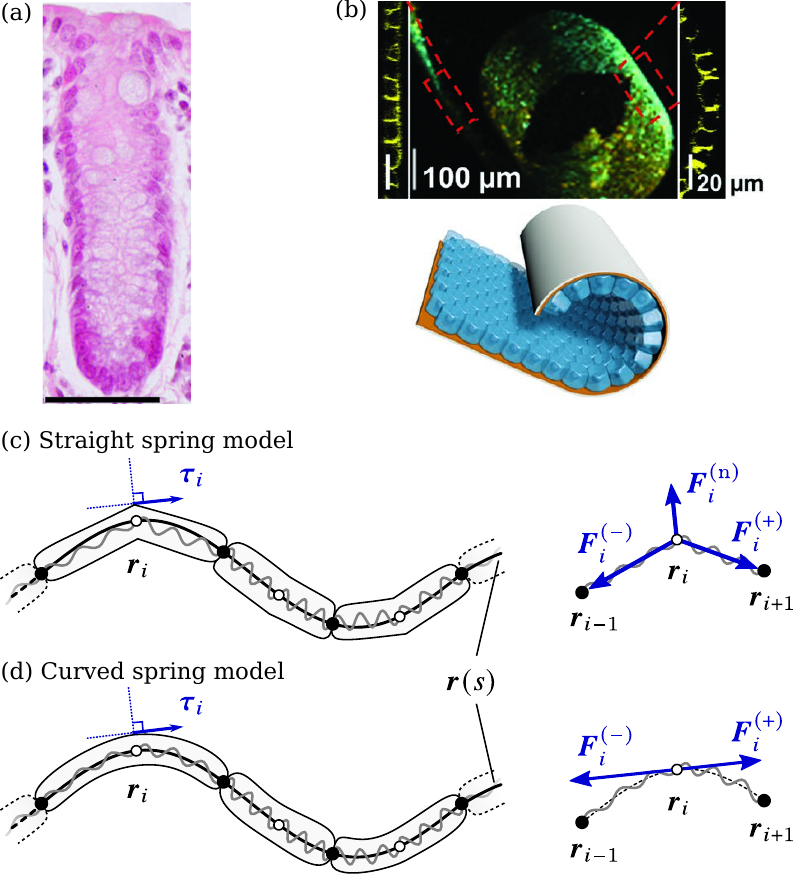}
    \caption{\cbeg (a) Colonic crypt showing a monolayer of epithelial cells in cross section (dark pink); scale bar=50\,$\muup$m. Reproduced from~\cite{dunn2012} under the terms of the Creative Commons Attribution License; (b) Epithelial cells on a substrate with transiently induced curvature. Adapted from~\cite{schamberger2023} under the terms of the CC-BY Creative Commons Attribution 4.0 license (\url{https://creativecommons.org/licenses/by/4.0/}); \cend (c) Straight spring model and (d) curved spring model of the mechanical interaction between cells along an interface $\b r(s)$ (solid black line). In this figure, each cell is composed of $m=2$ springs (gray coils). Nodes within a cell are shown as white circles, and nodes connecting two cells are shown as black circles. (c) The force diagram for the straight spring model shows that restoring forces are directed along the secant line between two nodes on the interface. The normal reaction force $\b F_i^{(\text{n})}$ ensures that the net force is parallel to the unit tangent vector $\b \tau_i$ at node $i$. (d) In the curved spring model, the restoring forces $\b F_i^{(\pm)}$ at node $i$ are already parallel to $\b \tau_i$ and so there is no normal reaction force $\b F_i^{(\text{n})}$.}\label{fig-spring-models}
\end{figure}

In the present work, we focus on developing a mathematical model of mechanical interactions within a constant population of cells arranged in a confluent layer on a curved interface in two-dimensional space. \cbeg This may represent curved epithelial tissues seen in cross-section, such as monolayers of epithelial cells in colonic crypts, epithelial cell sheets on engineered substrates with induced curvature (Fig.~\ref{fig-spring-models}), or epithelial-like osteoblast layers on bone tissues. However, \cend we do not take into account other cell behaviours than the mechanical interactions of the cells. We aim to characterise how the distribution of cells along the curved interface evolves in space and time based on specifying the cells' mechanical properties, such as their stiffness and resting length. \cbeg A key goal of this study is to understand how the curvature of the interface may influence the stress state of the cells and the dynamics of their mechanical relaxation.

Our approach considers both discrete, individual-based models of the mechanical behaviour of the cells, and their continuum limit. The continuum limit captures the collective evolution of the models by a partial differential equation for a continuous cell density field. The consideration of \cend discrete models and of their continuum limit is of strong interest in mathematical biology to bridge the cellular scale and the tissue scale. Discrete models allow us to represent the cellular entities that compose biological tissues and to encode cell behaviours directly. Continuum models capture average behaviour at the tissue scale more readily. However, without appropriate mathematical theories linking the cellular scale to the tissue scale, it can remain challenging to understand how cellular scale parameters relate to tissue scale behaviours, and to quantify these behaviours~\citep{chopard1998,li2022b,vandenheuvel2024}.

Many models of cellular dispersion are based on random walks~\citep{codling2008}. These types of models are appropriate when motile cells migrate through free space and their interactions represent crowding constraints~\citep{chopard1998,simpson2010,li2022a,li2022b}, but they are not well-suited to modelling mechanical interactions occurring within confluent layers of cells. Previous works have studied mechanical interactions between confluent cells arranged along a one-dimensional \cbeg flat substrate\cend~\citep{murray2009,murray2011,murray2012,fozard2010,murphy2019,murphy2020,murphy2021,tambyah2020}. These works find that when cells interact mechanically like overdamped Hookean springs with stiffness constant $k$ and drag coefficient $\eta$, the density of cells $q(x,t)$ at position $x$ along the line and time $t$ evolves, in an appropriate continuum limit, according to the nonlinear diffusion equation
\begin{align}\label{cell-density-1D}
    \pd{q}{t} = \pd{}{x}\left(\frac{k}{\eta}\frac{1}{q^2}\pd{q}{x}\right).
\end{align}
This partial differential equation represents the mechanical relaxation of the tissue as a \emph{fast diffusion} process with diffusivity $D(q) = k/(\eta q^2)$, in which low densities disperse rapidly~\citep{newman1980,king2003}. It belongs to a class of porous medium equations with a negative exponent $-2$ for the density dependence of the diffusivity, representing fast diffusion~\citep{vasquez2006}. Equation~\eqref{cell-density-1D} provides a direct relationship between mechanical properties of a cell body ($k$ and $\eta$) and the mechanics-driven evolution of a continuous cell density at the tissue scale. For non-Hookean restoring force laws, other nonlinear density dependences of the diffusivity hold~\citep{murray2012,baker2019}.

In this paper, we generalise Eq.~\eqref{cell-density-1D} to represent the evolution of cell density due to mechanical relaxation in curved epithelial layers. We base our mathematical model on a discrete model of cells interacting as a chain of mechanical springs similar to those proposed in the above works, except that we consider the springs to be oriented along a curved interface. This interface represents a substrate providing support for the cells, such as an epithelial basement membrane \cbeg (Figure~\ref{fig-spring-models}a)\cend, or other biological tissues supporting confluent cell layers, such as osteoblasts lying on bone tissue~\citep{martin1998}.

\cbeg To the best of our knowledge, no previous work has considered discrete models of mechanical interactions between cells lying on curved interfaces with their continuum limit. Since most biological tissues possess some degree of curvature, it is important to understand when and how curvature may influence the mechanical relaxation of cells. The generalisation of mechanical models of cells on flat interfaces to curved interfaces introduces new possibilities for representing cellular mechanics. We propose two different spring models of cellular mechanics: one in which the springs are straight and bridge across discrete positions on the continuous interface, and one in which the springs are curved and follow the curved interface. The mechanical forces generated by these two spring models are different in general. They depend on the geometry of the interface in different ways, but they coincide on flat interfaces. It is important to understand how much of this cell-level detail is retained at the tissue level by exploring the rate of convergence of the continuum limit of these models.

In addition to improving the biological relevance of spring models of epithelial tissues by considering curved interfaces, our work also provides a new continuum limit procedure, based on expansions in small spring lengths. Each cell is assumed to comprise $m$ springs and we let $m\to\infty$. In this way, the number of cells and cell density remain finite while the length of the springs goes to zero. We feel that this procedure is more intuitive and it involves less intermediate steps than the continuum limits proposed by~\cite{murray2009} and by~\cite{murphy2019}.  It ensures that cell numbers remain finite on a finite domain, which is more biologically relevant to represent confined tissues. Furthermore, our procedure allows us to provide formal justifications for cell parameter rescalings with $m$ that are necessary for converging to consistent dynamics. These parameter scalings were previously mentioned in~\citep{murray2009,murphy2019} without rigorous justification. We also provide for the first time the rate of convergence of the discrete models to the continuum limit as $m\to\infty$, which explains in particular why continuous densities approximate cell densities from the discrete model particularly well at the midpoint of the springs~\citep{murphy2019}. \cend We also investigate time scales of mechanical relaxation in the discrete models and the stability of homogenous densities depending on the spring restoring force laws~\citep{murray2012}. Finally, we calculate both the tangential and normal stresses within the cellular tissue to illustrate how surface tension induced by the spring forces generates curvature-dependent normal stress.

\section{Discrete model of mechanical cell interactions}
Following previous works~\citep{murray2009,murray2011,murray2012,fozard2010,baker2019,murphy2019,murphy2020,murphy2021,tambyah2020}, we consider mechanical interactions between adjacent cells, which we interpret as a model of the cross-section of an epithelial cell monolayer (Figure~\ref{fig-spring-models}). Each cell is in contact with a single cell on either side. While these previous works assumed cells to be constrained to the $x$-axis, here we consider that cells lie on a static interface in two-dimensional space. The interface is described by a general curve $\b r(s)=\big(x(s), y(s)\big)$ parametrised by arc length $s\in [0,L]$, where $L$ is the total length of the interface. The interface may be an open curve, or a closed loop with $\b r(0)=\b r(L)$. It can have arbitrary shape, except that we assume $\b r(s)$ to be differentiable, meaning that there are no sharp corners and a tangent vector exists everywhere. This assumption is physically realistic for epithelial layers, and mathematically convenient to describe the orientation of the cells and the direction of mechanical forces acting on them.

In the following, we first present two discrete models of mechanical interaction occurring between the cells on the interface, one based on straight springs, and one based on curved springs. We then show in Section~\ref{sec:continuum-limit} that in an appropriate continuum limit in which the number of springs goes to infinity, both models converge to the same nonlinear diffusion equation for the local cell density.

Both discrete models assume that the shape and mechanical behaviour of a biological cell can be represented by a chain of $m$ internal mechanical springs along the cell body. \cbeg Cells are connected directly at a single node, which balances the mechanical force of the last spring of one cell and the first spring of the other cell \cend (Figure~\ref{fig-spring-models}c,d). To simplify notation, throughout the paper we denote by $i=0,\ldots, M$ the index of a node connecting two springs irrespective of whether this node is internal to a cell, or at the boundary between two cells. We denote the total number of cells by~$N$, so that the total number of springs is $M=mN$. If the interface is a closed loop, the last spring is connected to the first spring, so that node $i=M$ is identified with node $i=0$ (there are only $M$ distinct nodes in this case). \cbeg We note here that our representation in which springs represent cell bodies is similar to that of~\cite{fozard2010} but different from that of~\cite{murray2009} in which springs represent the interaction between cells. Our choice of representing a cell body with multiple connected springs follows our previous work~\citep{murphy2019}. It allows us to better represent boundary conditions on cells at tissue boundaries, and to model the mechanics of cells of different types (e.g., cancer cells, non-cancer cells) with different stiffnesses or resting lengths~\citep{murphy2019}. In curved geometries, increasing the number of springs per cell $m$ also enables us to model cell shape.\cend

Cell body motion occurs in dissipative environments where inertial effects can be neglected. We therefore assume an overdamped regime, in which the position $\b r_i(t)$ of node $i$ at time $t$ evolves according to
\begin{align}\label{discrete-evo-0}
    \eta \td{\b r_i}{t} = \b F_i^{(-)} + \b F_i^{(+)} + \b F_i^{(\text{n})},
\end{align}
where $\eta$ is the viscous drag coefficient, and $\b F_i^{(-)}$ and $\b F_i^{(+)}$ are the mechanical forces exerted on node $i$ due to the springs connecting it to nodes $i-1$ and $i+1$, respectively. To ensure that the nodes remain on the interface at all times, an additional normal reaction force $\b F_i^{(\text{n})}$ may be required in curved geometries, which represents the interaction of the cells with the substrate (Figure~\ref{fig-spring-models}). This additional force is not required in models of epithelia on flat interfaces~\citep{murray2009,murphy2019}. We now detail how the forces $\b F_i^{(-)}$, $\b F_i^{(+)}$, and $\b F_i^{(\text{n})}$ are defined in the straight spring model, and in the curved spring model.

\subsection{Straight spring model}
In the straight spring model, the line of action of the spring restoring force is directed along the secant line connecting two nodes (Figure~\ref{fig-spring-models}a), i.e.,
\begin{align}
    \b F_i^{(-)} = - f\big(\|\b r_i-\b r_{i- 1}\|\big) \frac{\b r_i-\b r_{i- 1}}{\|\b r_i-\b r_{i-1}\|}, \quad \b F_i^{(+)} = f\big(\|\b r_{i+1}-\b r_i\|\big) \frac{\b r_{i+1}-\b r_i}{\|\b r_{i+1}-\b r_i\|},
\end{align}
where $f(\ell)$ is the restoring force law that depends on spring length $\ell$. The normal reaction force $\b F_i^{(\text{n})}$ in Eq.~\eqref{discrete-evo-0} is defined to be exactly opposite to the component of $\b F_i^{(-)}+\b F_i^{(+)}$ normal to the interface at node~$i$, so that the net force and velocity $\d\b r_i/\d t$ of node $i$ in Eq.~\eqref{discrete-evo-0} are tangent to the interface at all times. Denoting with $\b\tau_i$ the unit tangent vector to the interface at node~$i$ in the direction of increasing index~$i$, the position of node $i$ evolves in the straight spring model as
\begin{align}\label{straight-spring-evo-1}
      \eta \td{\b r_i}{t} = \left[ \left( f\big(\|\b r_{i+1}-\b r_i\|\big)\frac{\b r_{i+1}-\b r_i}{\|\b r_{i+1}-\b r_i\|}  - f\big(\|\b r_i-\b r_{i-1}\|\big)\frac{\b r_i-\b r_{i-1}}{\|\b r_i-\b r_{i-1}\|}\right)\vdot \b\tau_i \right]\b \tau_i.
\end{align}
Since the nodes remain on the interface at all times, we can reference their position along the parametric curve $\b r(s)$ by their arc length coordinate $s_i(t)$, such that
\begin{align}\label{pos-vs-arclength}
  \b r_i(t) = \b r\big(s_i(t)\big).
\end{align}
Using the fact that the unit tangent vector of the interface at $s$ is $\b\tau(s)=\d\b r(s)/\d s$ and that $\b\tau_i = \b\tau\big(s_i(t)\big)$, we obtain from Eq.~\eqref{pos-vs-arclength}, \begin{align}\label{velocity-vs-arclength}
	\td{\b r_i(t)}{t} = \td{\b  r\big(s_i(t)\big)}{s}\td{s_i(t)}{t} = \b \tau_i\td{s_i(t)}{t}.
\end{align}
Substituting these expressions for $\b r_i$ and $\d\b r_i/\d t$ in terms $s_i$ and $\d s_i/\d t$ in Eq.~\eqref{straight-spring-evo-1} provides the evolution of the nodes' arc length coordinates:
\begin{align}\label{straight-spring-evo-2}
    \eta \td{s_i}{t} = \left[f(\|\b r(s_{i+1})-\b r(s_i)\|)\frac{\b r(s_{i+1})-\b r(s_i)}{\|\b r(s_{i+1})-\b r(s_i)\|} - f(\|\b r(s_i)-\b r(s_{i-1})\|)\frac{\b r(s_i)-\b r(s_{i-1})}{\|\b r(s_i)-\b r(s_{i-1})\|}\right]\vdot\b\tau_i.
\end{align}
Since $\b r(s)$ is a known parametric curve, we can solve Eq.~\eqref{straight-spring-evo-2} instead of Eq.~\eqref{straight-spring-evo-1} to evolve the discrete model, with appropriate boundary conditions when $i=0$ and $i=M$ (see Section~\ref{sec:bc}).

\subsection{Curved spring model}
An alternative model of mechanical interactions is to consider curved springs that follow the curvature of the interface (Figure~\ref{fig-spring-models}b), so that the length of a spring is the arc length between two nodes, and the line of action of the restoring forces $\b F_i^{(\pm)}$ is directed along the unit tangent $\b\tau_i$ at node $i$, i.e.,
\begin{align}
    \b F_i^{(-)} = -f(s_i-s_{i-1})\b\tau_i, \qquad \b F_i^{(+)} = f(s_{i+1}-s_i)\b\tau_i,
\end{align}
where $s_i$ is the arc length coordinate of node $i$. Since these forces are tangent to the interface, $\b F_i^{(\text{n})}=\b 0$. In the curved spring model, the position of node $i$ therefore evolves from Eq.~\eqref{discrete-evo-0} as
\begin{align}
    \eta\td{\b r_i}{t} = \left[ f(s_{i+1}-s_i) - f(s_i-s_{i-1})\right]\b \tau_i.
\end{align}
Using Eq.~\eqref{velocity-vs-arclength}, the evolution of arc length coordinates in this model is given by
\begin{align}\label{curved-spring-evo-2}
    \eta\td{s_i}{t} = f(s_{i+1}-s_i) - f(s_i - s_{i-1}).
\end{align}
Mathematically, Eq.~\eqref{curved-spring-evo-2} is of the same form as the evolution of coordinates $x_i$ in a straight chain of springs along the $x$-axis~\citep{murray2009,murray2011,murray2012,murphy2019}, except that $x_i$ are replaced by arc length coordinates $s_i$. It is interesting to note that Eq.~\eqref{curved-spring-evo-2} is manifestly independent of the shape of the parametric curve $\b r(s)$, but Eq.~\eqref{straight-spring-evo-2} is not. When nodes interact mechanically via straight springs, the evolution of their node positions along the interface depends on the shape of the interface.

\subsection{Restoring force law}
In both the straight spring model and the curved spring model we may consider a general restoring force law $f(\ell)$ that depends on the evolving length $\ell$ of the spring. In our numerical simulations, we consider three different force laws, including a Hookean restoring force law linear in the elongation of the spring, and two nonlinear restoring force laws. The Hookean restoring force law is
\begin{align}\label{Hookean-restoring-force}
    f(\ell) = k(\ell-a),
\end{align}
where $a$ is the resting spring length and $k$ is the spring stiffness.

\begin{figure}
        \centering\includegraphics[scale=0.85]{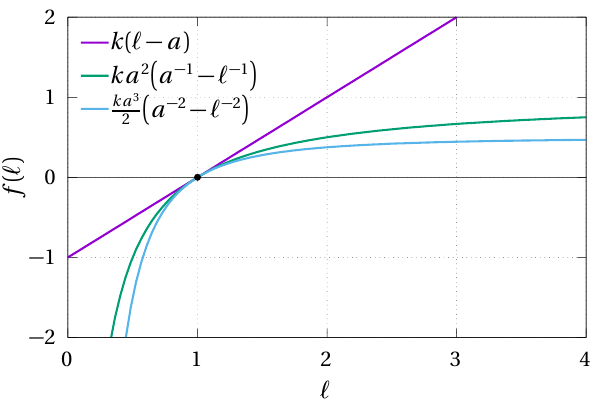}
        \caption{Comparison between the Hookean restoring force law $f(\ell)=k(\ell-a)$ (magenta), the nonlinear restoring force law $f(\ell)=ka^2(1/a-1/\ell)$ (green), and the nonlinear restoring force law $f(\ell)=(ka^3/2)\big(1/a^2-1/\ell^2\big)$ ($a=1$, $k=1$ in arbitrary units). The nonlinear restoring force scaling factors are such that their linearisation about the resting spring length $\ell=a$ gives the Hookean restoring force.}
        \label{fig-restoring-forces}
\end{figure}

Hooke's law means that cells can in principle be squeezed to zero length since the force remains finite when $\ell=0$. It also means that the further apart cells are, the stronger their mechanical interaction. These properties are biologically unrealistic for very small $\ell$ and for very large $\ell$, so we also consider nonlinear restoring force laws that diverge when $\ell\to 0$ and that remain bounded when the separation between cells increases ($\ell\to\infty$). Our specific choices of such restoring force laws are such that in the continuum limit, the mechanical relaxation of cells will be described by linear diffusion of cell density for our first choice, and by a common type of porous medium diffusion for our second choice (see Section~\ref{sec:cell-density}, Table~\ref{table:force-vs-diffusivity}). The first nonlinear restoring force law is
\begin{align}\label{nonlinear-restoring-force}
      f(\ell) = ka^2\left(\frac{1}{a}-\frac{1}{\ell}\right),
\end{align}
where $\rho_0=1/a$ is the resting spring density. The scaling factor $ka^2$ of this restoring force is such that the linearisation of Eq.~\eqref{nonlinear-restoring-force} about $\ell=a$ is the Hookean restoring force~\eqref{Hookean-restoring-force} with spring constant $k$ and resting length $a$ (Figure~\ref{fig-restoring-forces}).

The second nonlinear restoring force law is
\begin{align}\label{porous-restoring-force}
  f(\ell) = \frac{k a^3}{2}\left(\frac{1}{a^2} - \frac{1}{\ell^2}\right).
\end{align}
Likewise, the scaling factor $ka^3/2$ ensures that the linearisation of Eq.~\eqref{porous-restoring-force} about the resting length $a$ matches the Hookean restoring force (Figure~\ref{fig-restoring-forces}).

The restoring force laws $f(\ell)$ in Eqs~\eqref{Hookean-restoring-force}--\eqref{porous-restoring-force} are defined to be positive when the spring is elongated, and negative when the spring is compressed. We can define tangential stress along the cell body based on these force laws as follows. Considering first Hooke's law, the spring constant $k$ of a portion of cell body of resting length $a$ and cross-sectional area $A$ is $k=EA/a$, where $E$ is the cell body's Young modulus. The tangential stress along the cell body can thus be defined at each inner spring of a cell by
\begin{align}\label{tangential-stress}
    \sigma_{\tau\tau}(\ell) = - \frac{f(\ell)}{A} = - \frac{E}{k}\frac{f(\ell)}{a},
\end{align}
with the sign convention that compressive stress is positive and tensile stress is negative. For all the restoring forces that we consider, we will calculate tangential stress as $\sigma_{\tau\tau}/E$ using Eq.~\eqref{tangential-stress} as it only depends on $k$, $a$, and $\ell$ and it is always proportional to stress. The quantity $\sigma_{\tau\tau}/E$ represents strain for Hookean springs but not for nonlinear springs.

\subsection{Numerical simulations}\label{sec:bc}
Numerical simulations of the discrete model are based on evolving arc length coordinates using Eq.~\eqref{straight-spring-evo-2} for straight springs and Eq.~\eqref{curved-spring-evo-2} for curved springs, supplemented with the following boundary conditions. If the interface is an open curve, we fix the position of the boundary nodes $i=0$ and $i=M$ at the curve's end points, so that their arc length coordinates satisfy
\begin{align}\label{fixed-bc}
    s_0(t) = 0, \qquad s_M(t) = L, \qquad t\geq 0.
\end{align}
If the interface is a closed loop, we consider periodic boundary conditions on the node indices by identifying node $i=M$ with node $i=0$, so that
\begin{align}\label{pbc}
      s_0(t)=s_M(t), \qquad t\geq 0.
\end{align}

Our numerical simulations consider several interfaces for which we require an arc length parametrisation. Finding arc length parametrisations usually involves reparametrising an interface described by another, known parametrisation. For a circular interface of radius $R$ parametrised by an angle $\theta\in[0,2\pi)$, arc length is $s=R\theta$ so $\b r(s) = \big(R\cos(s/R), R\sin(s/R)\big)$, and the unit tangent vector is $\b\tau(s) = (-\sin(s/R), \cos(s/R))$. For other interface shapes, arc length $s$ is calculated from a known parametrisation $\widetilde{\b r}(u)$ of the interface by numerically integrating
\begin{align}
    s(u) = \int_0^u\!\!\!\d u\ \|\widetilde{\b r}'(u)\| \label{arclength}
\end{align}
using the composite Simpson's 1/3 rule. The arc length parametrisation of the interface is then obtained as $\b r(s)=\widetilde{\b r}\big(u(s)\big)$ where $u(s)$ is the inverse function of $s(u)$. This inverse function is estimated numerically by finding $u$ in Eq.~\eqref{arclength} such that $s(u)-s=0$ using the root-finding algorithm \texttt{findRoot} of \texttt{D}'s standard library, which is based on the \texttt{TOMS748} algorithm~\citep{dlang}. All our simulations assume arbitrary units of space and time. These units can be re-scaled to match any particular application by working with a relevant length scale defined by the cell diameter, and a mechanical relaxation time scale determined by the ratio $\eta/k$ (Section~\ref{sec:parameter-rescaling-discrete}).

Equations~\eqref{straight-spring-evo-2} and~\eqref{curved-spring-evo-2} are solved numerically using a simple explicit forward Euler time-stepping scheme with time step $\Deltaup t=0.001$, and with boundary conditions~\eqref{fixed-bc} for open curves, and~\eqref{pbc} for closed curves. Initial conditions assume that all cell boundaries are evenly spaced except for one that is offset by half a resting cell length. Spring boundaries are then initialised to be evenly spaced within a cell. For full detail on the algorithms used to solve the discrete models, the reader is referred to the \texttt{D} computer code available on \texttt{GitHub}~\citep{github-code}.

\begin{figure}[t]
	\centering\includegraphics[width=\textwidth]{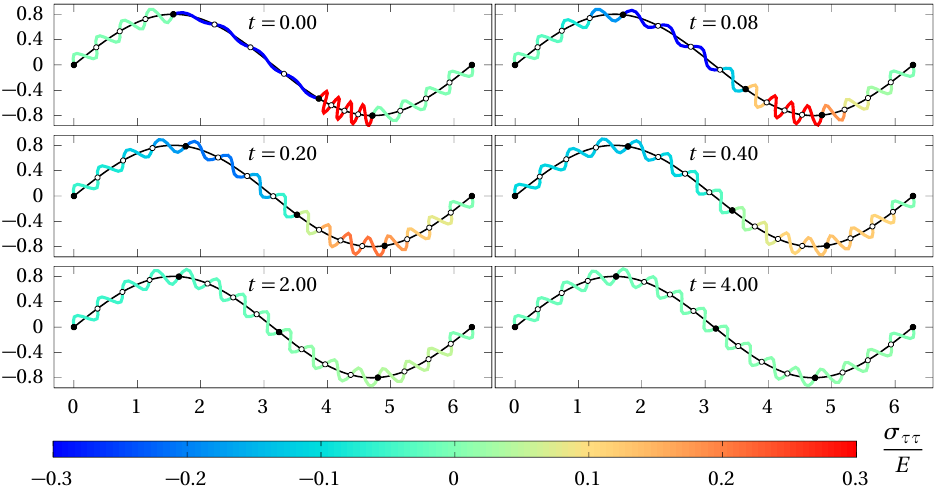}
    \caption{Time snapshots of the mechanical relaxation of $N=4$ cells with $m=4$ inner springs (stress-coloured coils) along the open curve $\widetilde{\b r}(u)=\big(u,R\sin(u)\big)$ (solid black curve) using the curved spring model and a Hookean restoring force. Cell boundaries are shown as black circles. Inner spring boundaries are shown as open circles. The resting spring length is chosen such that the steady state is stress-free; $R=0.8$, $k=4$, $\eta=0.25$, $a\approx 0.45$, $\Deltaup t = 0.001$.}\label{fig-sin}
\end{figure}
We start by considering a sinusoidal interface given by $\widetilde{\b r}(u)=\big(u, R\sin(u)\big)$ for $u\in[0,2\pi]$ (Figure~\ref{fig-sin}). The interface is populated with $N=4$ cells, each containing $m=4$ curved, Hookean springs, so that there are $M=16$ springs in total. The first and last spring boundaries are fixed at $\widetilde{\b r}(0)=(0,0)$ and $\widetilde{\b r}(2\pi)=(2\pi,0)$, respectively. The spring resting length is chosen as $a=L/M\approx 0.45$, where $L\approx 7.2$ is the length of the curve between $u=0$ and $u=2\pi$. Thus, springs are stress-free when in mechanical equilibrium. Since $m=4$, the resting cell length is $4a\approx 1.8$. In the initial configuration, the boundary between the second and third cell is displaced, such that the second cell is elongated by half a resting cell length, and the third cell is compressed by a half a resting cell length. Time snapshots of the numerical simulation are shown in Figure~\ref{fig-sin} where each spring $i$ of length $\ell_i=(s_i-s_{i-1})$ is coloured according to its normalised tangential stress $\sigma_{\tau\tau}(\ell_i)/E$. These snapshots show that the springs relax mechanically to even out mechanical stress by redistributing the position of their extremities along the curved interface.

\begin{figure}[t]
	\centering\includegraphics[width=\textwidth]{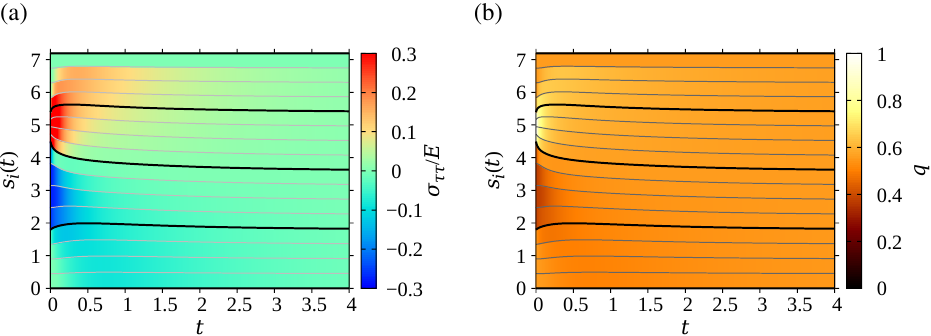}
    \caption{Evolution of spring boundary positions (thin gray lines) and cell boundary positions (thick black lines) in the simulations shown in Figure~\ref{fig-sin}. Each spring is coloured according to (a) its tangential stress $\sigma_{\tau\tau}(\ell_i)/E$; and (b) cell density~$q_i$.}
    \label{fig-sin-stress-density}
\end{figure}

The evolution of the arc length positions of the spring nodes for the same simulation is shown in Figure~\ref{fig-sin-stress-density} (solid black and grey lines). In Figure~\ref{fig-sin-stress-density}a, each spring $i$ is coloured according to its normalised tangential stress $\sigma_{\tau\tau}(\ell_i)/E$. In Figure~\ref{fig-sin-stress-density}b, each spring $i$ is coloured according to its local cell density $q_i=1/(m\ell_i)$. By evening out tangential stress along the interface with time, the mechanical forces between the cells also tend to even out cell density along the interface.

\section{Continuum limit}
\label{sec:continuum-limit}
In both the straight spring model and curved spring model, the cells relax to an equilibrium state through their mechanical interactions. The collective relaxation of the cells corresponds to the mechanical relaxation of the tissue. This tissue-scale relaxation can be described in a continuum limit by deriving the conservation equation that governs the evolution of cell density when the number of cells $N$ is kept constant and the number of springs per cell $m$ goes to infinity, $m\to\infty$~\citep{murphy2019}. This way of taking the continuum limit allows cell density to remain finite on a finite interface length $L$, while the total number of springs in the system goes to infinity, \cbeg $M=mN\to\infty$\cend.

We first show in Section~\ref{sec:equiv-straight-curved} that as $m\to\infty$ the dynamics of the straight spring model converges to that of the curved spring model. In Section~\ref{sec:cell-density}, we then propose an alternative and more intuitive derivation of the continuum limit of the evolution of cell density than the derivations provided in~\citet{murray2009,murphy2019,tambyah2020}. This new derivation enables us to find how the drag coefficient $\eta$, the spring stiffness $k$, and the resting length $a$ must be rescaled for the dynamics of the system to converge as the number of springs per cell increases. Furthermore, in Section~\ref{sec:parameter-rescaling-discrete} we propose an alternative justification for these parameter scalings based on analysing time scales of mechanical relaxation in the discrete models.

\subsection{Equivalence of straight and curved springs in the continuum limit}
\label{sec:equiv-straight-curved}
When the number of springs per cell $m$ goes to infinity, \cbeg the index $i$ runs over all natural numbers and \cend the distance between the nodes tends to zero, so that $\Deltaup s = s_{i+1}-s_{i}\to 0$. Using Eq.~\eqref{pos-vs-arclength} and the differentiability of $\b r(s)$, we therefore have
\begin{align}
  &\b r_{i+1} = \b r(s_i + \Deltaup s) = \b r_i + \b\tau_i\Deltaup s + \order(\Deltaup s), \notag
  \\&\b r_{i-1} = \b r(s_i-\Deltaup s) = \b r_i-\b\tau_i\Deltaup s + \order(\Deltaup s), \label{contlim-pos}
\end{align}
where $\order(\Deltaup s)$ terms are such that $\lim_{\Deltaup s\to 0}\order(\Deltaup s)/\Deltaup s=0$~\citep{olver1974}. Equations~\eqref{contlim-pos} show that the Euclidean distance between two nodes converges to the arc length distance along $\b r(s)$,
\begin{align}\label{contlim-arclength}
    \|\b r_{i+1}-\b r_i\| \sim \Deltaup s = s_{i+1}-s_i, \qquad m\to \infty,
\end{align}
and that the lines of force of $\b F_i^{(\pm)}$ in the straight spring model become parallel to the unit tangent vector $\b \tau_i$ at $\b r_i$:
\begin{align}\label{contlim-tangent}
      &\frac{\b r_{i+1}-\b r_{i}}{\|\b r_{i+1}-\b r_{i}\|} \to \b\tau_i, \qquad\frac{\b r_{i}-\b r_{i-1}}{\|\b r_{i}-\b r_{i-1}\|} \to \b\tau_i, \qquad m\to\infty.
\end{align}
Using Eqs~\eqref{contlim-arclength}--\eqref{contlim-tangent}, the evolution equation~\eqref{straight-spring-evo-2} for arc length coordinates in the straight spring model converges to the evolution equation~\eqref{curved-spring-evo-2} of the curved spring model as $m\to\infty$:
\begin{align}
  \eta\td{s_i}{t} &\sim \big(f(s_{i+1}-s_i)\b\tau_{i} - f(s_i-s_{i-1})\b\tau_i\big)\vdot\b\tau_i,\notag
  \\&= f(s_{i+1}-s_i)-f(s_i-s_{i-1}), \qquad m\to\infty. \label{discrete-evo}
\end{align}
Thus, for large number of springs per cell $m$, the evolution of spring boundaries for straight springs tends to that of curved springs.

\begin{figure}[t!]
	\centering\includegraphics[width=0.8\textwidth]{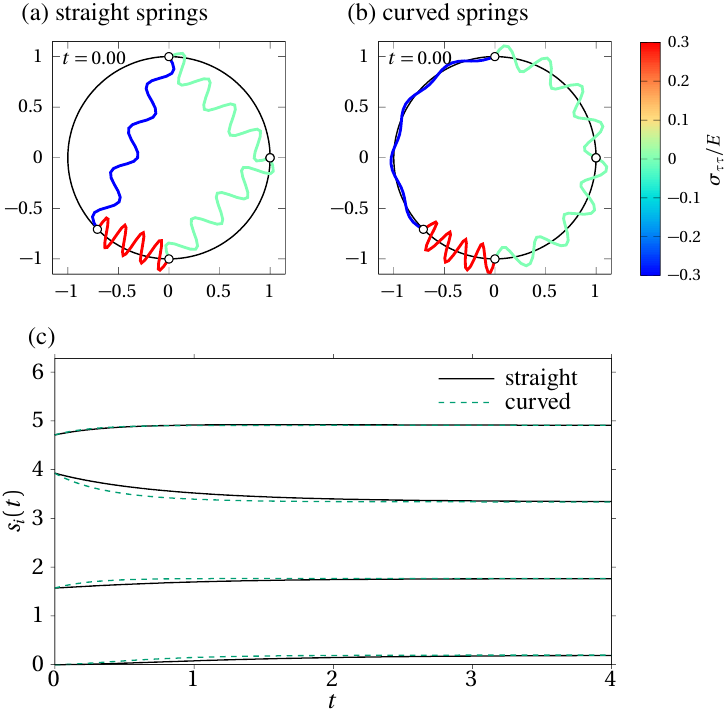}
    \caption{Evolution of spring boundary positions along the circle $\b r(s)=\big(R\cos(s/R),R\sin(s/R)\big)$ with straight and curved springs, $N=4$, $m=1$, $R=1$, $k=1$, $\eta=1$, $\Deltaup t=0.001$. (a) Initial condition with straight springs of resting length $a=2\sin(\pi/4) = \sqrt{2}$; (b) Initial condition with curved springs of resting length $a=2\pi/4$; (c) Comparison of the dynamics of mechanical relaxation between straight and curved springs.}
    \label{fig-circle}
\end{figure}
To illustrate the difference between the straight spring model and the curved spring model, we present numerical simulations in Figure~\ref{fig-circle} with few springs, $N=4$, $m=1$, and a circular interface. The spring resting length is chosen such that springs are stress-free when in mechanical equilibrium, meaning that $a=2\pi R/M$ for curved springs, and $a=2R\sin\big(\pi/M\big)$ for straight springs, which is the edge length of an $M$-sided regular polygon inscribed in a circle of radius $R$. When one of the nodes is offset by an arc length distance $a/2$ along the circle, resulting in one compressed cell adjacent to one extended cell, \cbeg the dynamics of mechanical relaxation is clearly different at short timescales between the straight spring model and the curved spring model (Figure~\ref{fig-circle}c). This difference emphasises the different ways that geometry influences these discrete models\cend. However, the mechanical equilibrium state is identical.

\begin{figure}[t!]
	\centering\includegraphics[width=0.8\textwidth]{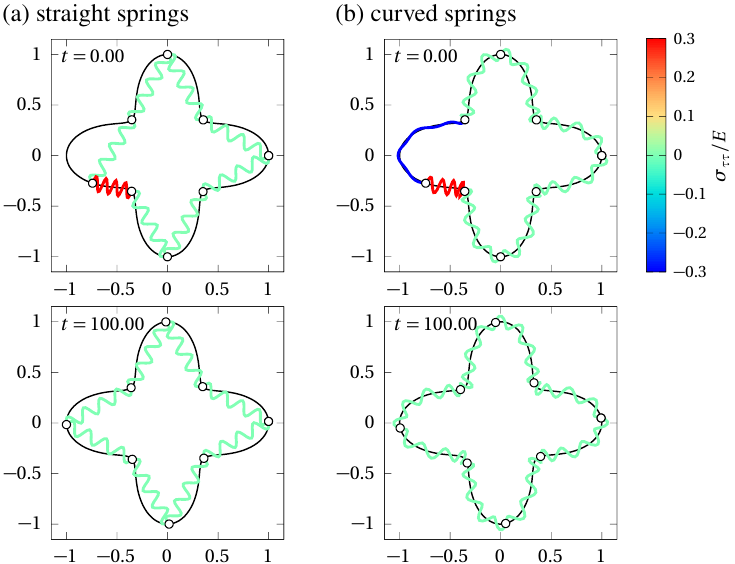}
    \caption{Comparison of mechanical relaxation between straight and curved springs on a cross-shaped interface with $N=8$ cells and $m=1$ spring per cell. The interface is defined in polar coordinates by the polar equation $R(\theta)=R_0\big(\cos^4(\theta) + \sin^4(\theta)\big)$;  $R_0=1$, $\eta=1$, $k=1$, $\Deltaup t=0.001$. The resting length $a$ is chosen such that there is no tangential stress in steady state. The snapshots show the initial configuration ($t=0$) and a mechanically relaxed configuration ($t=100$) for (a) straight springs with $a\approx 0.7368$; and (b) curved springs with $a\approx 0.8010$.}
    \label{fig-cross-snapshots}
\end{figure}
\begin{figure}[t!]
\centering\includegraphics[scale=0.95]{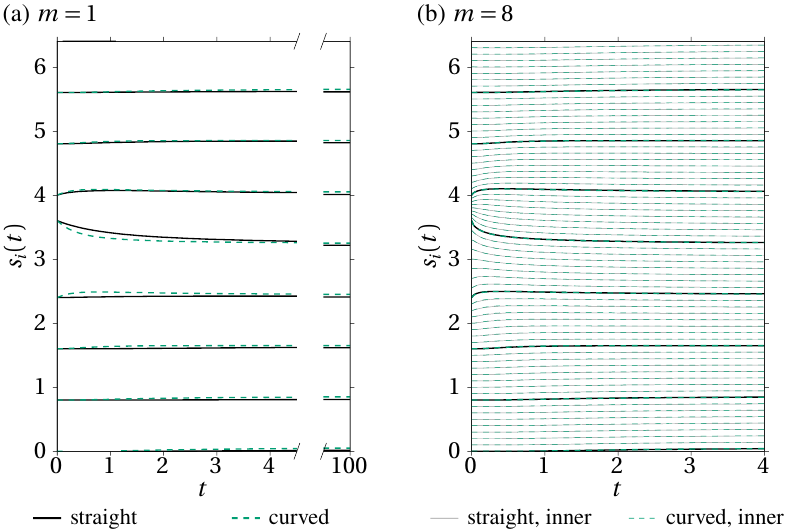}
\caption{Comparison of the evolution of spring boundary positions along the cross-shaped interface of Figure~\ref{fig-cross-snapshots} between straight springs and curved springs with $N=8$ cells. Cell boundaries are shown as solid black line (straight spring model) and thick dashed green lines (curved spring model). Inner spring boundaries within the cells are shown as thin grey lines (straight spring model) and thin dashed green lines (curved spring model). Spring resting lengths are chosen such that there is no tangential stress in steady state. (a) $m=1$ spring per cell, $k=1$, $\eta=1$, $a \approx 0.7368$ for straight springs, $a\approx 0.8010$ for curved springs, $\Deltaup t = 0.001$; (b) $m=8$ springs per cell, $k=8$, $\eta=1/8$, $a\approx 0.0995$ for straight springs, $a\approx 0.1001$ for curved springs, $\Deltaup t = 0.001$.}
\label{fig-cross-evo}
\end{figure}

Figure~\ref{fig-cross-snapshots} shows that if the interface is more complex, both the dynamics and the equilibrium state between straight springs and curved springs might not be the same. In the straight spring model, some nodes can remain stuck in regions of the interface with high curvature, due to the particular arrangement of spring force directions (Figure~\ref{fig-cross-snapshots}a). In contrast, spring force directions are always tangent to the interface with curved springs, so nodes can always slide away from regions with high curvature (Figure~\ref{fig-cross-snapshots}b). These differences between the straight spring and curved spring models quickly disappear as $m$ is increased \cbeg if the interface is smooth\cend. Figure~\ref{fig-cross-evo} compares the arc length positions of the spring nodes when $m=1$ and when $m=8$ in both spring models. It shows that both the dynamics and steady state of mechanical relaxation on the cross-shaped boundary of Figure~\ref{fig-cross-snapshots} become indistinguishable already for $m=8$ springs per cell (Figure~\ref{fig-cross-evo}b). However, to obtain similar dynamics and relaxation times of the spring nodes when $m=1$ and when $m=8$, it is necessary to rescale the spring resting length by a factor $1/m$, and the ratio $k/\eta$ by a factor $m^2$, which we do by rescaling $\eta$ by $1/m$ and $k$ by $m$. While these scalings were already proposed by \citet{murray2009,murphy2019}, there were not previously justified mathematically. We will justify these scalings rigorously in Sections~\ref{sec:cell-density} and~\ref{sec:parameter-rescaling-discrete}. \cbeg We also note that if the interface is not smooth, nodes in the straight spring model could remain stuck at cusps of the interface even in the continuum limit, so that there can be differences between the straight spring model and the curved spring model even in the continuum limit.\cend
\clearpage

\subsection{Evolution of cell density in the continuum limit}
\label{sec:cell-density}
Equation~\eqref{discrete-evo} is identical to the evolution of node coordinates $x_i$ in a straight chain of springs along the $x$ axis~\citep{murray2009,murphy2019}, with the only difference that the coordinates are now arc lengths $s_i$ along a curved interface. The geometry of the interface does not appear explicitly in this equation. The evolution of cell density in the continuum limit is therefore identical to that found in~\citep{murray2009,murphy2019} subject to substituting $x$ for arc length~$s$. Murray et al.~\citep{murray2009,murray2011,murray2012} and Tambyah et al.~\citep{tambyah2020} derive the continuum limit based on considering density as the reciprocal of the metric $\p x_i/\p i$ for a continuous spring index $i$. Murphy et al.~\citep{murphy2019,murphy2020} derive the same continuum limit based on local spatial averages. We provide an alternative derivation, based on a conceptually simpler expansion of cell density in small spring lengths. This new derivation has the benefit of (i)~clearly justifying scalings required on parameters of the discrete model to reach consistent dynamics as the number of springs increases; (ii)~calculating the order of approximation of the discrete model provided by the continuum limit. This order of approximation justifies in particular why in numerical simulations the continuous density matches the discrete density near the midpoint between spring nodes, which was previously observed, but not explained~\citep{murphy2019}.

Since the straight spring model becomes identical to the curved spring model when $m\to\infty$, we consider the curved spring model and denote by $\ell_i = (s_i-s_{i-1})$ the length of the $i^\text{th}$ spring. The local spring density in the discrete model is given by $\rho_i = 1/\ell_i$.
The evolution of spring density is calculated from Eq.~\eqref{curved-spring-evo-2} as:
\begin{align}
    \td{\rho_i}{t} &= -\frac{1}{\ell_i^2}\td{\ell_i}{t} = -\frac{1}{\ell_i^2}\left(\td{s_i}{t}-\td{s_{i-1}}{t}\right)\notag
  \\&= -\frac{1}{\eta\ell_i^2}\Big( f(s_{i+1}-s_i) - f(s_i-s_{i-1}) - \big(f(s_i-s_{i-1}) - f(s_{i-1}-s_{i-2})\big)\Big)\notag
\\&= -\frac{1}{\eta}\frac{f(\ell_{i+1}) - 2f(\ell_i) + f(\ell_{i-1})}{\ell_i^2}.\label{evo-spring-density-discrete}
\end{align}
Equation~\eqref{evo-spring-density-discrete} is exact for curved springs, and asymptotic for straight springs when $m\to\infty$ via Eq.~\eqref{discrete-evo}. To take the continuum limit, we now introduce a continuous function $\rho(s,t)$ of arc length position~$s$ and time~$t$, such that
\begin{align}\label{rho-cont}
    \rho\big(\overline{s}_i, t\big) = \rho_i(t) = 1/\ell_i(t),
\end{align}
where
\begin{align}\label{midpoint-arclength}
	\overline{s}_i  = \frac{1}{2}\big(s_i + s_{i-1}\big)
\end{align}
is the (time-dependent) arc length midpoint between the spring boundaries $s_{i-1}$ and $s_i$. We also define the restoring force law as a function of spring density instead of length by introducing
\begin{align}\label{f-tilde}
\widetilde{f}(\rho) = f(1/\rho).
\end{align}
Because spring lengths tend to zero, spring density diverges as $\Order(m)$ as $m\to\infty$. Since each cell contains $m$ springs, we also introduce a continuous, local cell density function $q(s,t)$ of arc length position $s$ and time $t$ based on $\rho(s,t)$, defined as
\begin{align}\label{q}
     q(s,t) = \frac{\rho(s,t)}{m}.
\end{align}
We thus expect cell density $q(s,t)$ to remain finite, of order $\Order(1)$, in the continuum limit. However, we will see that additional requirements on the restoring force and drag coefficient are necessary for the evolution of $q(s,t)$ in the continuum limit to match the evolution of the discrete models with $m \gg 1$. We first explore the large-$m$ behaviour of Eq.~\eqref{evo-spring-density-discrete} for the (diverging) spring density $\rho(s,t)$, and then consider under what conditions the evolution of cell density $q(s,t)$ is well defined in the limit $m\to\infty$.

Using the continuous function $\rho(s,t)$, Eq.~\eqref{rho-cont}, and the fact that $\ell_i = (s_{i}-s_{i-1}) = 1/\rho(\overline{s}_i,t)$, Eq.~\eqref{evo-spring-density-discrete} can be rewritten as
\begin{align}\label{evo-density-midpoint}
  \td{}{t}\rho(\overline{s}_i,t)  &\sim -\frac{1}{\eta}\frac{\widetilde{f}\big(\rho(\overline{s}_{i+1},t)\big)-2\widetilde{f}\big(\rho(\overline{s}_i,t)\big) + \widetilde{f}\big(\rho(\overline{s}_{i-1},t)\big)}{\ell_i^2}, \qquad m\to\infty.
\end{align}
The right hand side is similar to a discretised second order derivative of $\widetilde{f}\big(\rho(s,t)\big)$ with respect to arc length evaluated at $\overline{s}_i$. However, $\overline{s}_{i-1}$, $\overline{s}_i$, and $\overline{s}_{i+1}$ are not evenly spaced, and the time derivative in the left hand side must also take into account that $\overline{s}_i$ is time dependent. In Appendix~\ref{appx:continuum-limit}, we show that by properly expanding $\overline{s}_{i+1}$ and $\overline{s}_{i-1}$ about $\overline{s}_i$ in the limit $m\to\infty$ and accounting for the time dependence of $\overline{s}_i$ in the left hand side, Eq.~\eqref{evo-density-midpoint} asymptotically becomes the partial differential equation
\begin{align}
  \pd{\rho}{t}(\overline{s}_i,t) &= -\frac{1}{\eta}\pd{^2}{s^2}\widetilde{f}\big(\rho(\overline{s}_i,t)\big) \left(1+ \Order\left(\frac{1}{m^2}\right)\right),\quad m\to\infty.\label{evo-spring-density-continuous}
\end{align}
It is important to emphasise that the $\Order(1/m^2)$ correction relies on matching the continuum and discrete spring densities at the arc length midpoint of a spring in Eq.~\eqref{rho-cont}. Matching at other points along the spring decreases accuracy to $\Order(1/m)$ (Appendix~\ref{appx:continuum-limit}). As expected from these results, the arc length midpoint of a spring is precisely where numerical simulations show excellent agreement between discrete densities and continuum densities (Section~\ref{sec:numerical-simulations-cont}). The order of accuracy of the continuum limit was not known in previous works.

The evolution of cell density can now be obtained from Eq.~\eqref{evo-spring-density-continuous} \cbeg by substituting $\rho(\overline{s}_i,t)=m q(\overline{s}_i,t)$ and by substituting $\overline{s}_i$ for an arbitrary, continuous arc length coordinate $s$\cend:
\begin{align}
     \pd{q}{t}(s, t) = -\pd{^2}{s^2}\left(\frac{1}{\eta m}\widetilde{f}\big(m q(s,t)\big)\right) + \Order\left(\frac{1}{m^2}\right) \label{evo-cell-density-1}.
\end{align}
For this partial differential equation to be well defined as $m\to\infty$, the restoring force and drag coefficient must be such that the following limit exists:
\begin{align}\label{restoring-force-contlim}
    \lim_{m\to\infty}\frac{1}{\eta m}\widetilde{f}\big(m q(s,t)\big) = F\big(q(s,t)\big),
\end{align}
where the limit defines a cell-density-dependent restoring force function $F(q)$. If this limiting force function $F(q)$ is reached with order $\Order(1/m^2)$, then the evolution equation for cell density in the continuum limit is a conservation equation with flux given by the gradient of $F$ with respect to arc length~$s$, $\p F\big(q(s,t))/\p s$, up to second-order corrections in $1/m$:
\begin{align}\label{evo-cell-density-2}
  \pd{q}{t}(s,t) = -\pd{^2}{s^2}F\big(q(s,t)\big) + \Order
 \left(\frac{1}{m^2}\right).
\end{align}
We can recast the limiting equation as the nonlinear diffusion equation
\begin{align}\label{nonlinear-diffusion}
    \pd{q}{t}(s,t) = \pd{}{s}\left(D\big(q(s,t)\big)\pd{q}{s}(s,t)\right), \qquad \text{where}\quad D(q) = - \td{F}{q}(q).
\end{align}
Equation~\eqref{nonlinear-diffusion} was derived previously for arbitrary restoring force laws by \citet{murray2012} but for the diverging density of springs \cbeg(Eq.~\eqref{evo-spring-density-continuous})\cend, and therefore without providing rigorous conditions on the required scalings of the restoring force law with $m$ exhibited in Eq.~\eqref{restoring-force-contlim}. Equation~\eqref{nonlinear-diffusion} was derived by \citet{murphy2019} for the cell density, which is well-defined in the continuum limit, when the restoring force is Hookean, but without providing rigorous justifications for the restoring force law scaling and the order of approximation.

Solutions to the partial differential equations~\eqref{evo-cell-density-2} or~\eqref{nonlinear-diffusion} require the specification of initial and boundary conditions, which are chosen to match those of the discrete model. When the interface is an open curve, fixing the position of node $i=0$ and $i=M$ in the discrete model in Eqs~\eqref{fixed-bc} means that no spring can move past the fixed boundary nodes $\b r_0(t) = \b r(0)$ and $\b r_M(t) = \b r(L)$. Accordingly, no-flux boundary conditions on $q(s,t)$ are imposed at $s=0$ and $s=L$\cbeg~\citep{murray2009,fozard2010,baker2019,murphy2019,tambyah2020}\cend:
\begin{align}\label{no-flux-bc-q}
	\pd{q}{s}(0,t)= \pd{q}{s}(L,t)=0, \quad t>0.
\end{align}
When the interface is a closed loop, periodic boundary conditions on the node indices and their arc length positions in the discrete model in Eqs~\eqref{pbc} translate as periodic boundary conditions on~$q(s,t)$:
\begin{align}\label{pbc-q}
	q(0,t) = q(L,t), \quad t>0.
\end{align}

\paragraph{Hookean restoring force} In~\citet{murphy2019}, numerical simulations of the discrete models were performed by rescaling the parameters $a, k, \eta$ in Eq.~\eqref{Hookean-restoring-force} with $m$. For clarity, we now add a superscript `$(m)$' to these quantities. The scalings were defined as follows:
\begin{align}
  &a^{(m)} =\frac{a^\ast}{m}, \qquad k^{(m)} = k^\ast m,
  \qquad \eta^{(m)} = \frac{\eta^\ast}{m}, \label{rescale-a-k-eta}
\end{align}
where $a^\ast$, $k^\ast$, and $\eta^\ast$ are $\Order(1)$ as $m\to\infty$ and can be considered mechanical properties of the cell as a whole. Similar parameter rescalings are also mentioned in Murray et al.~\citep{murray2009} without justification. These rescalings ensure that simulations of the discrete model with linear restoring force match simulations of the continuum cell density. The scaling of the resting spring length $a^{(m)}$ is suggested by springs being smaller with increasing~$m$. The scaling of the spring constant $k^{(m)}$ is suggested by $k^\ast=k^{(m)}/m$ being the equivalent spring constant for $m$ identical Hookean springs in series. The reduction of the drag coefficient $\eta^{(m)}$ with $m$ can be argued based on needing to retain a constant total drag force on the cell as its number of springs increases. Equation~\eqref{restoring-force-contlim} provides a mathematical justification for these scalings as they ensure that the limit is well defined. Indeed, with the Hookean restoring force law in Eq.~\eqref{Hookean-restoring-force}, we have
\begin{align}
    F(q(s,t)) = \lim_{m\to\infty}\frac{k^{(m)}}{m^2\eta^{(m)}}\left(\frac{1}{q(s,t)} - ma^{(m)}\right)= \frac{k^\ast}{\eta^\ast}\left(\frac{1}{q(s,t)} - a^\ast\right).\label{F}
\end{align}
The limit in Eq.~\eqref{F} defines $\eta^\ast F$ as the restoring force law of a single cell with cell resting length $a^\ast$ and cell spring constant $k^\ast$, where $\eta^\ast$ is the drag coefficient of the cell. Substituting the expression for $F(q)$ obtained from Eq.~\eqref{F} into Eq.~\eqref{evo-cell-density-2} shows that cell density relaxes mechanically along the curved interface according to the nonlinear diffusion equation
\begin{align}\label{evo-cell-density-Hooke}
    \pd{q}{t} = \pd{}{s}\left(D(q)\pd{q}{s}\right), \qquad D(q)=\frac{k^\ast}{\eta^\ast}\frac{1}{q^2}.
\end{align}
Equation~\eqref{evo-cell-density-Hooke} is the same as that found for cells arranged along the $x$-axis except it involves the arc length coordinate $s$ instead of $x$, see Eq.~\eqref{cell-density-1D}~\citep{murray2009,murray2012,murphy2019}. We also note that with the scalings in Eqs~\eqref{rescale-a-k-eta}, the limit in Eq.~\eqref{F} is reached for any value of $m$, so the order of approximation of Eq.~\eqref{evo-cell-density-Hooke} is indeed~$\Order(1/m^2)$.

Tangential stress $\sigma_{\tau\tau}$ in the cell body, given by Eq.~\eqref{tangential-stress}, also converges to a well-defined limit as $m\to\infty$, since the length $\ell$ of a spring scales as $\Order(m^{-1})$:
\begin{align}
    \sigma_{\tau\tau} = E\left(\frac{\ell}{a^{(m)}}-1\right) \longrightarrow E\left(\frac{1/q}{a^\ast}-1\right), \qquad m\to\infty.
\end{align}

\paragraph{Nonlinear restoring forces}
With the nonlinear restoring force~\eqref{nonlinear-restoring-force} and the parameter scalings in Eqs~\eqref{rescale-a-k-eta}, the limit~\eqref{restoring-force-contlim} becomes
\begin{align}\label{nonlinear-restoring-force-cell}
    F\big(q(s,t)\big) = \lim_{m\to\infty}\frac{1}{m\eta^{(m)}} k^{(m)}\big(a^{(m)}\big)^2 \big(\rho_0-mq(s,t)\big) = D_0\big(q_0-q(s,t)\big),
\end{align}
where $D_0=k^\ast (a^\ast)^2/\eta^\ast$ is a diffusion constant, and $q_0=\rho_0/m=1/a^\ast$ is the resting cell density. Substituting the expression for $F(q)$ from Eq.~\eqref{nonlinear-restoring-force-cell} into Eq.~\eqref{evo-cell-density-2} shows that cell density relaxes mechanically according to linear diffusion along the curved interface:
\begin{align}
    \pd{q}{t} = D_0\pd{^2q}{s^2},
  \label{evo-cell-density-lineardiff}
\end{align}
where $\p^2/\p s^2$ corresponds to the one-dimensional Laplace--Beltrami operator along the curve $\b r(s)$~\citep{berger2003}. With the nonlinear restoring force~\eqref{nonlinear-restoring-force}, tangential stress in the cell body likewise converges to a finite value as $m\to\infty$ since spring length $\ell$ is $\Order(m^{-1})$:
\begin{align}\label{stress}
    \sigma_{\tau\tau} = E\left(1 - \frac{a^{(m)}}{\ell}\right) \longrightarrow E\left(1-\frac{a^\ast}{1/q}\right), \qquad m\to\infty.
\end{align}
If the spring restoring force in Eq.~\eqref{nonlinear-restoring-force} is written with an arbitrary scaling factor $\alpha^{(m)}$ such that $f(\ell)=\alpha^{(m)}\big(1/a^{(m)}-1/\ell\big)$, then $\alpha^{(m)}$ must scale as $\Order(1/m)$ for the continuum limit in Eq.~\eqref{restoring-force-contlim} to be well defined. In this case, the diffusion constant is given by $D_0=\alpha^\ast/\eta^\ast$ where $\alpha^\ast=\lim_{m\to\infty}m\alpha^{(m)}$.

With the nonlinear restoring force~\eqref{porous-restoring-force} and the scalings in Eqs~\eqref{rescale-a-k-eta}, the limit~\eqref{restoring-force-contlim} becomes
\begin{align}
  F\big(q(s,t)\big) = \lim_{m\to\infty}\frac{1}{m\eta^{(m)}} \frac{k^{(m)}\big(a^{(m)}\big)^3}{2} \big(\rho_0^2-m^2q^2(s,t)\big) = \frac{D_0^\ast}{2}\big(q_0^2-q^2(s,t)\big),
\end{align}
where $D_0^\ast = k^\ast(a^\ast)^3/\eta^\ast$. Cell density therefore relaxes mechanically according to a well-known type of porous medium diffusivity:
\begin{align}
  \pd{q}{t} = \pd{}{s}\left(D(q)\pd{q}{s}\right), \qquad  D(q) = D_0^\ast\, q.
  \label{evo-cell-density-porousdiff}
\end{align}
Equation~\eqref{evo-cell-density-porousdiff} is the porous medium equation with exponent one for the density dependence of diffusivity, also known as the Boussinesq equation~\citep{boussinesq1904,vasquez2006}.
Here too, tangential stress converges to a finite value as $m\to\infty$:
\begin{align}
	\sigma_{\tau\tau} = \frac{E}{2} \left(1-\left(\frac{a^{(m)}}{\ell}\right)^2\right) \longrightarrow \frac{E}{2}\left( 1- \left(\frac{a^\ast}{1/q}\right)^2\right), \qquad m\to\infty.
\end{align}
If the spring restoring force in Eq.~\eqref{porous-restoring-force} is written with an arbitrary scaling factor $\beta^{(m)}$ such that $f(\ell) = \beta^{(m)}\big(1/(a^{(m)})^2 - 1/\ell^2\big)$, then $\beta^{(m)}$ must scale as $\Order(1/m^2)$ for the continuum limit in Eq.~\eqref{restoring-force-contlim} to be well defined. In this case, $D_0^\ast = 2 \beta^\ast/\eta^\ast$ where $\beta^\ast=\lim_{m\to\infty}m^2\beta^{(m)}$. 

Table~\ref{table:force-vs-diffusivity} summarises the relationship between the restoring force laws assumed for the springs, and the corresponding diffusivity of cell density found in the continuum limit. Other nonlinear spring restoring forces can be devised to give a wanted diffusivity function $D(q)$ in the continuum limit. The procedure to find the spring restoring force is to first integrate $F'(q)=-D(q)$, and set the integration constant such that $F(1/a^\ast)=0$. The cell restoring force $\eta^\ast F(q)$ can then be rescaled to a spring restoring force $f(\ell)$ by making the parameters of $\eta^\ast F(q)$ dependent on $m$ based on substituting $a^\ast=m a^{(m)}$ and on utilising the limit~\eqref{restoring-force-contlim}. Alternatively, linearising $\eta^\ast F(q)$ about $q_0=1/a^\ast$ and matching it to Hooke's restoring force law of the cell in Eq.~\eqref{F} will provide scalings for the parameters of $\eta^\ast F(q)$ with $m$ via those known for Hooke's law in Eqs~\eqref{rescale-a-k-eta}.

\begin{table*}
	\caption{Relationship between spring restoring force law $f(\ell)$ in the discrete model and diffusivity of cell density $D(q)$ in the continuum limit. All cases assume the spring resting length scaling $a^{(m)} = a^\ast/m$ and drag coefficient scaling $\eta^{(m)} = \eta^\ast/m$ from Eqs~\eqref{rescale-a-k-eta}.}\label{table:force-vs-diffusivity}
	\begin{center}\begin{tabular}{ccc}
  \toprule
  $f(\ell)$ & Scaling ($m\to\infty$) & $D(q)$
  \\\midrule
	$k^{(m)}(\ell-a^{(m)})$ & $k^{(m)}\sim k^\ast m$ & $(k^\ast/\eta^\ast)\ q^{-2}$
	\\$\alpha^{(m)}\big((a^{(m)})^{-1}-\ell^{-1}\big)$ & $\alpha^{(m)}\sim \alpha^\ast/m$ & $(\alpha^\ast/\eta^\ast)$
	\\$\beta^{(m)}\big((a^{(m)})^{-2}-\ell^{-2}\big)$ & $\beta^{(m)}\sim \beta^\ast/m^2$ & $(2\beta^\ast/\eta^\ast)\ q$
  \\\bottomrule
	\end{tabular}\end{center}
\end{table*}

\subsection{Numerical simulations}\label{sec:numerical-simulations-cont}
The partial differential equation~\eqref{nonlinear-diffusion} with boundary conditions~\eqref{no-flux-bc-q} or~\eqref{pbc-q} is discretised in space using the method of lines with central finite difference approximations of both partial derivatives involving $s$, and a uniform discretisation of the arc length parameter in $[0,L]$ with $1000$ intervals of length $\deltaup s = L/1000$~\citep{lynch2005}. The resulting system of ordinary differential equations is then solved by the Tsitouras 5/4 Runge--Kutta method using the \texttt{Tsit5} method of the Julia package \texttt{DifferentialEquations}~\citep{julia}. For more detail on the algorithms used to solve the continuum models, the reader is referred to the \texttt{Julia} computer code available on \texttt{GitHub}~\citep{github-code}.

\begin{figure}[t!]
	\centering\includegraphics[width=\textwidth]{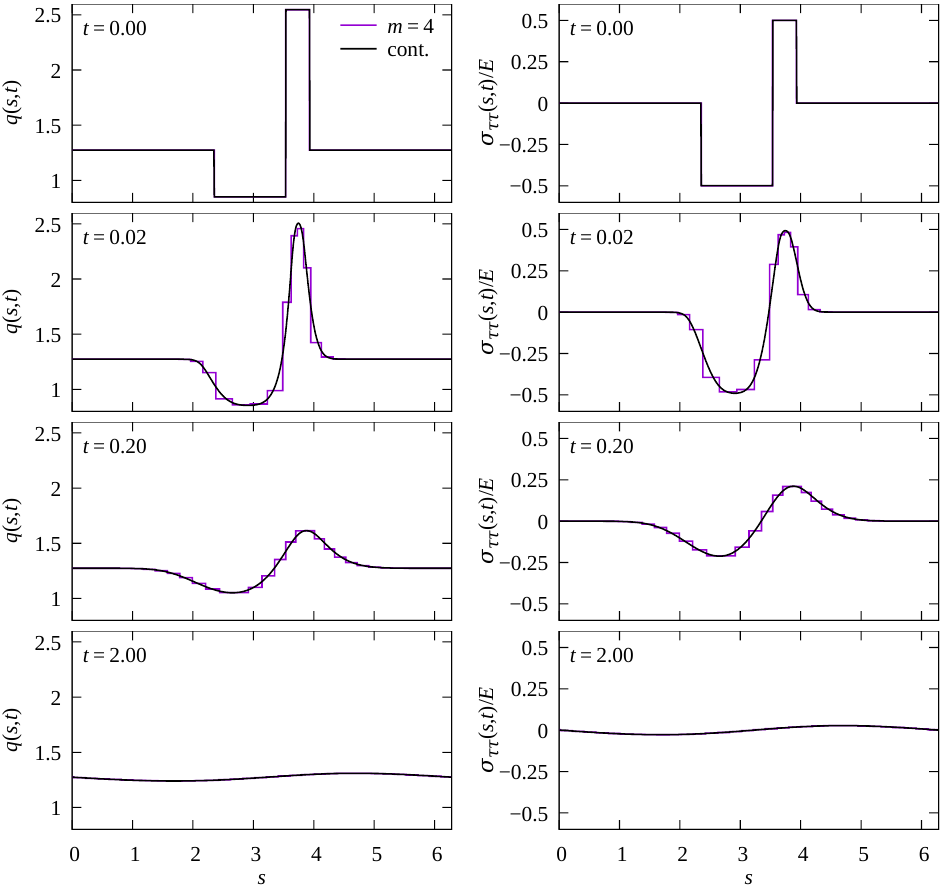}
	\caption{Comparison of density and stress state between discrete model simulations (magenta) and continuum model simulations (black) for $N=8$ cells around the unit circle with $m=4$ at times $t=0, 0.02, 0.2, 2$ (curved Hookean springs); $R=1$, $k^\ast=1$, $\eta^\ast=1$, $\Deltaup t=0.001$. The initial condition considers that one cell boundary is displaced along the circle by half a resting cell length $a^\ast=2\pi/8$ such that one cell is initially stretched 50\% ($\sigma_{\tau\tau}/E=-0.5$), and its neighbouring cell is compressed 50\% ($\sigma_{\tau\tau}/E=0.5$), like in Figure~\ref{fig-circle}b.}\label{fig-disc-cont}
\end{figure}
Figure~\ref{fig-disc-cont} compares numerical simulations of the continuum partial differential equation~\eqref{nonlinear-diffusion} with discrete model simulations performed with $N=8$ cells and $m=4$ curved Hookean springs per cell along a circular interface. The evolution of the continuous cell density in Eq.~\eqref{nonlinear-diffusion} and the evolution of the discrete cell density in Eq.~\eqref{evo-spring-density-discrete} are independent of interface geometry, so the choice of a circular interface is unimportant, except for using periodic boundary conditions. Figure~\ref{fig-disc-cont} shows that there is an excellent match between the models for the cell density profiles and for the tangential stress profiles at all times \cbeg when compared at the midpoint of the discrete springs\cend, even for a total number of springs as low as~32. The continuous solution curves cross the stepwise discrete densities and tangential stress near the midpoint between spring boundaries, which is consistent with our continuum limit derivation suggesting $\Order(1/m^2)$ accuracy \cbeg there. However, for other points between the spring boundaries, the continuum limit is not a good approximation of the discrete model for small number of springs. Accuracy for these points is only $\Order(1/m)$.  Since the curvature of the interface does not affect the dynamics in the continuum limit, these results are similar to those of \cite{murphy2019} obtained for springs confined to the $x$ axis, except \cbeg that they involve arc length coordinates and \cend that periodic boundary conditions are employed here.

\begin{figure}[t]
	\includegraphics[width=\textwidth]{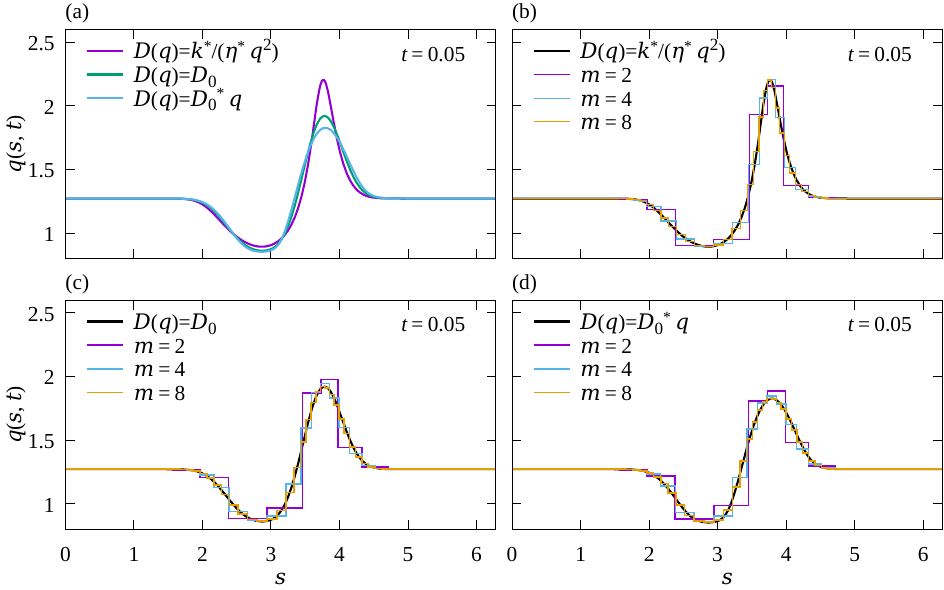}
	\caption{Comparison of cell density relaxation with different restoring forces. The discrete model is solved on a circular interface with curved springs for $N=8$ cells and $m=2,4,8$ springs per cells; $R=1$, $k^\ast=1$, $\eta^\ast=1$, $a^\ast=2\pi/8$, $\Deltaup t = 0.001$. The initial condition is the same for all cases and matches that of Figure~\ref{fig-disc-cont} ($t=0$) with one elongated cell adjacent to one compressed cell. Density profile along the interface are shown at time $t=0.05$. (a) Continuum cell density profiles obtained by solving Eq.~\eqref{nonlinear-diffusion} with $D(q)=k^\ast/(\eta^\ast q^2)$, $D(q)=D_0=k^\ast(a^\ast)^2/\eta^\ast\approx 0.62$, and  $D(q)=D_0^\ast q$, where $D_0^\ast=k^\ast(a^\ast)^3/\eta^\ast\approx 0.48$; (b)--(d) Comparison between discrete and continuum model simulations for (b) Hookean springs leading to $D(q)=k^\ast/(\eta^\ast q^2)$; (b) nonlinear springs leading to linear diffusion $D(q)=D_0$; (c) nonlinear springs leading to porous medium diffusion $D(q)=D_0^\ast q$.}
	\label{fig-disc-cont-many-forces}
\end{figure}
In Figure~\ref{fig-disc-cont-many-forces}, we show how the choice of restoring force law affects the dynamics of mechanical relaxation. Figure~\ref{fig-disc-cont-many-forces}a compares the continuous cell densities profiles $q(s,t)$ at time $t=0.05$ obtained by solving Eq.~\eqref{nonlinear-diffusion} with the diffusivities obtained in the continuum limit for the three choices of restoring force laws in Eqs~\eqref{Hookean-restoring-force}--\eqref{porous-restoring-force} (see Table~\ref{table:force-vs-diffusivity}). The different diffusivities result in slightly different cell density profiles. Figures~\ref{fig-disc-cont-many-forces}b--d show for each choice of restoring force law how quickly increasing the number of springs per cell makes the discrete cell densities converge to the continuum cell density. While the discrete density at the arc length midpoint of a \cbeg spring \cend is closest to the continuum cell density, it is clear that any other choice of point along the spring will also converge to the continuum density as $m\to\infty$. It is interesting to point out here that the discrete model simulations can be used as an alternative, robust and conservative discretisation method for a broad class of diffusion equations with linear and nonlinear diffusivities.

\subsection{Discrete-model justification of parameter rescaling}\label{sec:parameter-rescaling-discrete}
The scaling with $m$ of the restoring force law in Eq.~\eqref{restoring-force-contlim}, and the corresponding scalings with $m$ of its parameters in Eqs~\eqref{rescale-a-k-eta}, were shown in Section~\ref{sec:continuum-limit} to be required for the continuum limit to define a partial differential evolution equation for cell density. In this section, we justify the scalings~\eqref{rescale-a-k-eta} by analysing the time scales of mechanical relaxation in the evolution equations~\eqref{discrete-evo} of the discrete model. For clarity of the presentation, we omit momentarily the superscripts `$(m)$' in $k$, $\eta$, and $a$ in this section.

\begin{figure}
        \centering\includegraphics[scale=0.9]{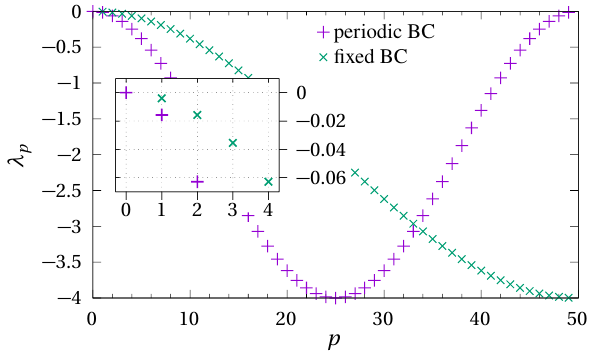}
        \caption{Eigenvalues $\lambda_p$ of the matrix $B$ corresponding to periodic boundary conditions, $p=0,\ldots,M-1$ (plus signs), and eigenvalues $\lambda_p$ of the matrix $\mathcal{B}$ corresponding to fixed boundary conditions, $p=1,\ldots, M-1$ (cross signs), with $M=50$ springs. The inset shows a close-up view of the first eigenvalues. The zero eigenvalue of the matrix $B$ for periodic boundary conditions corresponds to the steady state, see Eq.~\eqref{discrete-sol}. For fixed boundary conditions, the steady state is not an eigenvector of the matrix $\mathcal{B}$, see Eq.~\eqref{discrete-sol-fixed}.}
        \label{fig-eigenvalues}
\end{figure}

\paragraph{Hookean springs with periodic boundaries} Let $\b s(t)=\big(s_0(t),\ldots,s_{M-1}(t)\big)$ be the vector of arc length positions of the $M$ distinct nodes. With the convention that $s_M(t)=s_0(t)$ due to the periodic boundary condition, and assuming a linear restoring force (Hookean springs), Eqs~\eqref{discrete-evo} can be written in matrix form as
\begin{align}\label{discrete-evo-matrix}
  \td{}{t}\b s = \frac{k}{\eta}B \b s,  \qquad B=\pmat{-2 & 1 & 0 & \cdots & 0 & 1
  \\ 1 & \ddots & \ddots & \ddots &  & 0
  \\ 0 & \ddots & \ddots & \ddots & \ddots & \vdots
\\\vdots & \ddots & \ddots & \ddots & \ddots & 0
  \\ 0 &  & \ddots & \ddots & \ddots & 1
    \\ 1 & 0 & \cdots & 0 & 1 & -2
  }.
\end{align}
The matrix $B$ is an $M \times M$ circulant matrix. It is symmetric and therefore has real eigenvalues $\lambda_p$, $p=0,\ldots,M-1$. Decomposing the initial condition in the basis formed by the eigenvectors $\b v^{(p)}$ associated with $\lambda_p$, we can write $\b s(0)=\sum_{p=0}^{M-1}a_p\b v^{(p)}$. The solution to Eq.~\eqref{discrete-evo-matrix} is then given by
\begin{align}\label{discrete-sol}
    \b s(t) = \sum_{p=0}^{M-1} a_p \exp\left(\frac{k}{\eta}\lambda_p t\right)\b v^{(p)}.
\end{align}
To find the characteristic times of the evolution of $\b s(t)$ we now calculate the eigenvalues $\lambda_p$ by solving the eigenvalue problem $B\b v=\lambda \b v$ with $\b v=(v_0,\ldots v_{M-1})$. Given the matrix $B$ in Eq.~\eqref{discrete-evo-matrix}, the components of $\b v$ must satisfy
\begin{align}\label{eigenvalue-problem}
    v_{n+1} - (\lambda+2)v_n + v_{n-1} = 0, \qquad n=0,\ldots, M-1,
\end{align}
with the convention that $v_M=v_0$ and $v_{-1}=v_{M-1}$. For $\lambda=0$ the eigenvector $\b v^{(0)}$ solution of Eq.~\eqref{eigenvalue-problem} has components $v^{(0)}_n = n$, $n=0,\ldots, M-1$. In Eq.~\eqref{discrete-sol}, this corresponds to a steady-state contribution $p=0$ in which all the nodes are evenly spaced. To find the other eigenvalues, we take advantage of the periodicity of the system and expand each component $v_n$ of $\b v$ in the Fourier basis $\e^{\i p\,2\pi n/M}$, $p=0,\ldots,M-1$:
\begin{align}\label{fourier}
    v_n = \sum_{p=0}^{M-1} \widehat{v}_p\,\e^{\i p\frac{2\pi n}{M}},
\end{align}
so that $v_n = v_{n+M}$ for all $n$. Substituting Eq.~\eqref{fourier} into Eq.~\eqref{eigenvalue-problem} gives
\begin{align}
    \sum_{p=0}^{M-1} \widehat{v}_p\,\e^{\i p\frac{2\pi n}{M}} \left(\e^{\i p\frac{2\pi}{M}} - (\lambda+2) + \e^{-\i p\frac{2\pi}{M}}\right) = 0, \qquad n=0,\ldots,M-1.
\end{align}
For this equality to be satisfied for all $n$ we must have
\begin{align*}
    \e^{\i p\frac{2\pi}{M}} - (\lambda + 2) +\e^{-\i p \frac{2\pi}{M}} = 0,\qquad p=0,\ldots,M-1,
\end{align*}
so that
\begin{align}
\lambda_p = -2 + 2 \cos\left(\frac{2\pi p}{M}\right), \qquad p=0,\ldots,M-1.
\end{align}
The largest eigenvalue is $\lambda_0=0$, as found before. All the other eigenvalues are negative, so that the solution \eqref{discrete-sol} always converges to the steady state
\begin{align}
    \b{\overline s} = \lim_{t\to\infty}\b s(t) = a_0\b v^{(0)}
\end{align}
in which all nodes are equally spaced. In the long-time limit, the rate of convergence to steady state is dominated by the largest nonzero eigenvalues $\lambda_1=\lambda_{M-1}<0$ (Figure~\ref{fig-eigenvalues}). In the continuum limit where the number of springs $M\to\infty$, we have
\begin{align*}
    \lambda_1 = \lambda_{M-1} =-2+2\cos\left(\frac{2\pi}{M}\right) \sim - \frac{4\pi^2}{M^2}, \qquad M\to \infty,
\end{align*}
so that
\begin{align}
    \b s(t) \sim \b{\overline s}  + \big(a_1\b v^\ast + a_{M-1}\b v^{(M-1)}\big) \exp\left(-4\frac{k}{\eta}\frac{\pi^2}{M^2} t\right), \qquad t\to\infty, M\to\infty.
\end{align}
It is clear from this long-time asymptotic behaviour that the dynamics of mechanical relaxation becomes independent of $M$ in the limit $M\to\infty$ if
\begin{align}\label{asymptotic-params}
    \frac{k}{\eta} = \Order\big(M^2\big), \qquad M\to\infty.
\end{align}
This argument can repeated for other pairs of eigenmodes with faster relaxation rates ($\lambda_2=\lambda_{M-2}$, etc). We conclude that the dynamics of the discrete model becomes independent of $M$ in the continuum limit $M\to\infty$, provided that the parameters $k=k^{(m)}$ and $\eta=\eta^{(m)}$ scale with $M$ according to Eq.~\eqref{asymptotic-params}, which is satisfied by the scalings in Eqs~\eqref{rescale-a-k-eta}. The characteristic time scale of relaxation to steady state in this case is given by
\begin{align}\label{relaxation-time-pbc}
    T_\text{relax} \sim \frac{1}{4}\frac{\eta^{(m)}M^2}{k^{(m)}\pi^2} = \frac{1}{4}\frac{\eta^\ast N^2}{k^\ast\pi^2}, \qquad m\to\infty.
\end{align}

\paragraph{Hookean springs with fixed boundaries} With fixed boundaries, a formal justification of the scalings in Eq.~\eqref{rescale-a-k-eta} can also be done. The $M$ spring system is such that the arc length position of the first node is fixed at $s_0(t)=s_0=0$, but is now distinct from the arc length position of the last node held fixed at $s_M(t)=s_M=L$.

Let $\b s(t)=\big(s_1(t),\ldots,s_{M-1}(t)\big)$ be the vector of arc length positions of the $M-1$ interior nodes, with the convention that $s_i(t) < s_{i+1}(t)$ for $t \ge 0$ and $i = 0, \ldots, M-1$. With a linear restoring force (Hookean springs), Eqs~\eqref{discrete-evo} can be written in matrix form as
\begin{align}\label{discrete-evo-matrix-fixed}
  \td{}{t}\b s = \frac{k}{\eta}\mathcal{B} \b s + \dfrac{k}{\eta}\mathcal{A},  \qquad     \mathcal{B}=\pmat{-2 & 1 & 0 & \cdots  & 0
  \\ 1 & \ddots & \ddots & \ddots &  \vdots
\\0 & \ddots & \ddots & \ddots &  0
  \\ \vdots & \ddots & \ddots & \ddots &  1
    \\ 0 & \cdots & 0 & 1 & -2
  }
 ,\qquad
  \mathcal{A} = \pmat[c]{s_0\\
                      0 \\
                      \vdots \\
                      0 \\
                      s_M
  } = \pmat[c]{0\\0\\\vdots \\ 0\\ L}.
\end{align}
Here, $\mathcal{B}$ is an $(M-1)\times (M-1)$ Toeplitz tridiagonal matrix and $\mathcal{A}$ is a constant vector of length $M-1$. Writing $\b s(0)=\sum_{p=1}^{M-1}a_p\b v^{(p)}$ in the basis formed by the eigenvectors $\b v^{(p)}$ associated with the (real) $M-1$ eigenvalues $\lambda_p$ of $\mathcal{B}$, the solution of Eq.~\eqref{discrete-evo-matrix-fixed} is
 \begin{align}\label{discrete-sol-fixed}
    \b s(t) = \b{\overline s} + \sum_{p=1}^{M-2} a_p \exp\left(\frac{k}{\eta}\lambda_p t\right)\b v^{(p)},
\end{align}
where $\b{ \overline s}$ is a constant vector of length $M-1$ whose $i$th element is given by $i L/M$, for $i=1,\ldots,M-1$, which corresponds to the steady state with equally spaced nodes. In this case the eigenvalues of $\mathcal{B}$ can be written as \citep{kouachi2006}
\begin{align}
\lambda_p = -2 + 2\cos\left(\frac{\pi p}{M}\right), \qquad p=1,\ldots,M-1.
\end{align}
Thus, $\lambda_p < 0$ for all $p=1,\ldots M-1$ (Figure~\ref{fig-eigenvalues}) and all components of the time-dependent term in Eq.~\eqref{discrete-sol-fixed} decay to zero as $t \to \infty$.  The rate of approach to the long-time limit~$\b{\overline s}$ is dominated by the largest eigenvalue
\begin{align*}
    \lambda_1 = -2+2\cos\left(\frac{\pi}{M}\right) \sim - \frac{\pi^2}{M^2}, \qquad M\to \infty,
\end{align*}
so that the decay to the steady state is proportional to $\exp\left(- k \pi^2 t /(\eta M^2)\right)$ as $t \to \infty$ and $M \to \infty$.  For the dynamics of mechanical relaxation to become independent of $M$ as $M \to \infty$ we once again must require  $k/\eta = \Order(M^2)$ as in Eq.~\eqref{asymptotic-params}. The characteristic time scale of relaxation with fixed boundaries is
\begin{align}
    T_\text{relax}\sim \frac{\eta^{(m)}M^2}{k^{(m)}\pi^2} = \frac{\eta^\ast N^2}{k^\ast\pi^2},
\end{align}
i.e., the dynamics is four times slower than with periodic boundaries (compare with Eq.~\eqref{relaxation-time-pbc}).

\paragraph{Nonlinear restoring forces}
Similar arguments can be made for the long-time relaxation dynamics with nonlinear restoring forces. In this case, Eqs~\eqref{discrete-evo} form a nonlinear system of ordinary differential equations that do not have a closed-form solution. To analyse the long-term dynamics of this nonlinear system, we can linearise Eqs~\eqref{discrete-evo} about the mechanical equilibrium state $\b{\overline s}=(\overline s_0,\ldots, \overline s_{M-1})$ in which all the nodes are equally spaced: $\overline s_i = iL/M$, where $L$ is the length of the interface. The deviation to steady state $\xi_i = s_i-\overline s_i$ evolves in the linear approximation regime according to
\begin{align}\label{discrete-evo-nonlinear}
      \eta\td{\xi_i}{t} = k\big(\xi_{i+1}-2\xi_i+\xi_{i-1}\big),
\end{align}
where $k$ is now defined as $k=f'(L/M)$. Equation~\eqref{discrete-evo-nonlinear} can be recast in the same matrix form as Eq.~\eqref{discrete-evo-matrix} with periodic boundary conditions, or as Eq.~\eqref{discrete-evo-matrix-fixed} with fixed boundary conditions, leading to the time evolutions in Eqs~\eqref{discrete-sol} and~\eqref{discrete-sol-fixed} for $\xi_i$, respectively. It is clear from the expressions of these time evolutions that if $k>0$ then the state with evenly spaced nodes is a linearly stable steady state, and the dynamics of the discrete model near the steady state is independent of $M$ in the continuum limit $M\to\infty$ provided that the parameters $k$ and $\eta$ scale according to Eq.~\eqref{asymptotic-params}. If $k<0$, then the state with evenly spaced nodes is unstable. This situation occurs for example when the restoring force law is chosen to derive from the Lennard--Jones potential, which accounts for aggregation and negative diffusion for a restricted range of density. This restoring force law was found by \citet{murray2012} to result in evolutions in which small clusters of varying densities emerge since the state with evenly spaced nodes is unstable. In this situation, the linear approximation in Eq.~\eqref{discrete-evo-nonlinear} quickly fails and the continuum limit does not represent the discrete model dynamics well~\citep{murray2012}. In all the restoring forces we consider in this paper, $k>0$. For the nonlinear restoring force $f(\ell) = \alpha^{(m)} (\rho_0-1/\ell)$ with $\rho_0=1/a^{(m)}$, we have
\begin{align}
    k^{(m)} = f'\left(\frac{L}{M}\right) = \alpha^{(m)}\frac{M^2}{L^2},
\end{align}
so that with the scalings~\eqref{rescale-a-k-eta}, the parameter $\alpha^{(m)}$ must scale as
\begin{align}
	\alpha^{(m)} = k^{(m)}\frac{L^2}{M^2} = \Order\left(\frac{1}{m}\right).
\end{align}
For the nonlinear restoring force $f(\ell)=\beta^{(m)}(\rho_0^2 - 1/\ell^2)$, we have
\begin{align}
	k^{(m)} = f'\left(\frac{L}{M}\right) = \beta^{(m)}\frac{M^3}{L^3},
\end{align}
so that the parameter $\beta^{(m)}$ must scale as
\begin{align}
	\beta^{(m)} = k^{(m)}\frac{L^3}{M^3} = \Order\left(\frac{1}{m^2}\right).
\end{align}
These scalings are also summarised in Table~\ref{table:force-vs-diffusivity}.

\section{Surface tension and normal stress}
\begin{figure}[t!]
	\centering\includegraphics[scale=0.76]{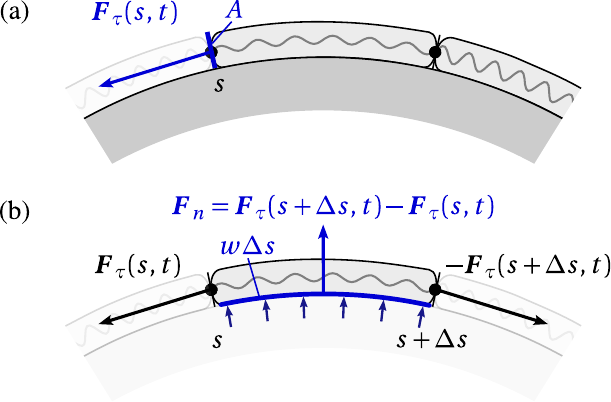}
	\caption{Tangential and normal stresses. (a) Tangential stress $\sigma_{\tau\tau}$ is defined as the tangential component of the inner force $\b F_\tau$ (blue arrow) divided by the cross-sectional surface area $A$; (b) Normal stress $\sigma_{nn}$ is defined as the normal component of the inner force $\b F_n$ (large blue arrow) divided by the contact surface area $w\Deltaup s$, where $w$ is the cell width in the out-of-plane direction. The normal force $\b F_n$ is the net reaction force exerted by substrate on the cell between the arc length positions $s$ and $s+\Deltaup s$ (small blue arrows). This normal force is induced by the tangential forces on curved portions of the interface only (see text for further detail).}
	\label{fig-surface-tension}
\end{figure}

Confluent cells on the curved substrate $\b r(s)$ experience two stress components: the tangential stress $\sigma_{\tau\tau}$ introduced in Eq.~\eqref{tangential-stress} due to mechanical interactions with neighbouring cells, and a normal stress $\sigma_{nn}$ due to normal reaction forces exerted by the substrate onto the cell (Figure~\ref{fig-surface-tension}). Tangential stress is due to the tangential inner force
\begin{align}
  \b F_\tau(s,t) = - \widetilde{f}\left(\rho(s,t)\right)\b \tau(s)
\end{align}
exerting within the cellular layer (Figure~\ref{fig-surface-tension}a), i.e.,
\begin{align}\label{tangential-stress2}
	\sigma_{\tau\tau} = \frac{F_\tau^\tau}{A} = \frac{\b \tau\vdot\b F_\tau}{A} = - \frac{\widetilde{f}(\rho)}{A} = -\frac{E}{k}\frac{f(\ell)}{a},
\end{align}
as in Eq.~\eqref{tangential-stress}, where $F_\tau^\tau=\b\tau\vdot\b F_\tau$ is the tangential component of $\b F_\tau$, $A$ is the cross-sectional area over which this force is exerted, and our sign convention is such that tensile stress is negative and compressive stress is positive.

The normal reaction forces exerting on the cellular layer are induced by the tangential forces on curved portions of the interface only (Figure~\ref{fig-surface-tension}b). \cbeg These normal reaction forces are present \cend even in the continuum limit when the normal force $\b F^{(\text{n})}$ in Eq.~\eqref{discrete-evo-0} acting on spring boundary nodes in the straight spring model goes to zero. Normal reaction forces acting on the cell body are also present for curved springs which assume no normal force $\b F^{(\text{n})}$ on spring boundaries. The normal stress $\sigma_{nn}$ is generated by the fact that an elongated cell body senses some of the curvature of the interface, such that the tangential forces exerted at the cell boundaries will have different directions and generate a net force in the normal direction (Figure~\ref{fig-surface-tension}b). The normal reaction force $\b F_n$ of the substrate is opposite to this net force, and distributed over the contact area between the cell and the substrate. Assuming a cell width $w$ in the out-of-plane direction, the surface area of the cellular layer in contact with the substrate over an arc length $\Deltaup s$ is $w \Deltaup s$, so that
\begin{align}
	\sigma_{nn} = \frac{F_n^n}{w\Deltaup s} = \frac{\b n\vdot\b F_n}{w\Deltaup s} = \frac{\b n\vdot\big(\b F_\tau(s+\Deltaup s, t) - \b F_\tau(s,t)\big)}{w\Deltaup s}.
\end{align}
To assign a local, continuous normal stress $\sigma_{nn}(s,t)$ at the arc length position $s$ of the interface, we take $\Deltaup s\to 0$ and also assume the continuum limit $m\to\infty$. In these limits, the net mechanical force per unit length of interface exerting on a portion of the interface of length $\Deltaup s$ is given by
\begin{align}
  \lim_{\Deltaup s\to 0}\frac{\b F_\tau(s+\Deltaup s, t)-\b F_\tau(s, t)}{\Deltaup s} &=  \pd{}{s}\b F_\tau(s,t) = - \pd{}{s}\left\{\widetilde{f}\big(\rho(s,t)\big)\right\}\b \tau(s) + \widetilde{f}\big(\rho(s,t)\big)\kappa(s)\b n(s),
\end{align}
where we have used the fact that the curvature of the interface $\kappa(s)$ at the arc length position $s$ is defined such that $\b\tau'(s) = - \kappa(s)\b n(s)$~\citep{berger2003}. Our curvature sign convention is such that $\kappa <0$ where the subtrate is concave, and $\kappa>0$ where the substrate is convex. Therefore,
\begin{align}\label{normal-stress}
	\sigma_{nn} = \gamma(s,t)\kappa(s),
\end{align}
where
\begin{align}\label{surface-tension}
	 \gamma(s,t) = \frac{\widetilde{f}\big(\rho(s,t)\big)}{w}  
\end{align}
is by definition the surface tension of the cellular layer. Equation \eqref{normal-stress} is similar to the Young--Laplace equation of surface tension of droplets. Because normal stress $\sigma_{nn}$ is induced by tangential forces, $\sigma_{nn}$ is nonzero only if the tangential forces are nonzero and if the interface has some curvature. For example, if the substrate is convex and cells are stretched, $\kappa>0$ and $\widetilde{f}>0$, so that $\sigma_{nn}>0$, which represents compressive normal stress. In contrast, tangential stress $\sigma_{\tau\tau}$ in Eq.~\eqref{tangential-stress2} is manifestly independent of the geometry of the interface, since the evolution of cell density governed by Eq.~\eqref{nonlinear-diffusion} is also independent of the shape and curvature of the interface.
\begin{figure}[t!]
	\centering\includegraphics[width=\textwidth]{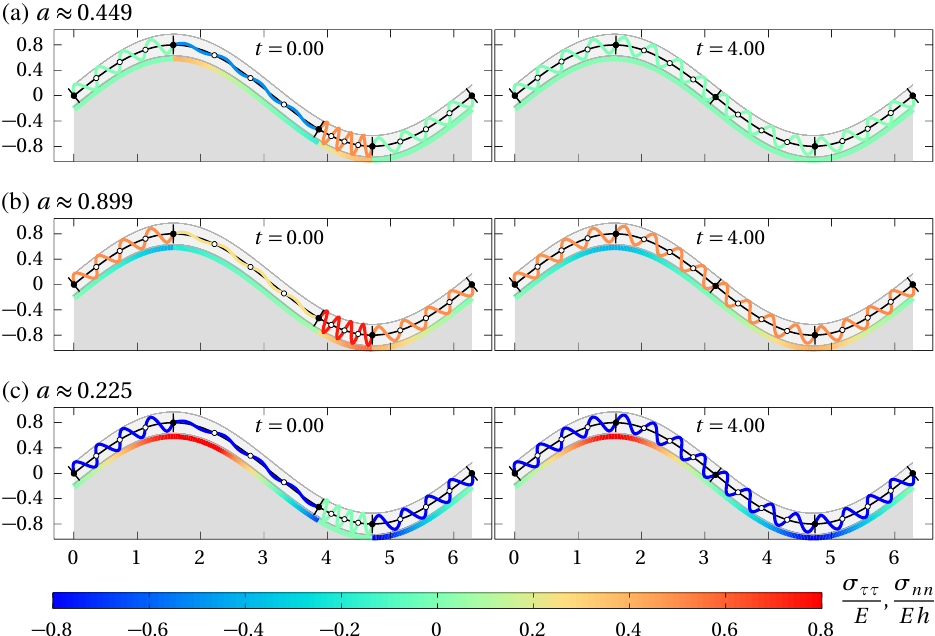}	
    \caption{Tangential and normal stresses along the open curve $\widetilde{\b r}(u)=\big(u,R\sin(u)\big)$ (solid black curve) using the curved spring model with $N=4, m=4$. Springs are coloured by the tangential stress $\sigma_{\tau\tau}/E$. The contact interface between cells and the substrate is coloured by the normal stress $\sigma_{nn}/(Eh)$. Simulation parameters are as in Figure~\ref{fig-sin}, i.e., $R=0.8$, $k=4$, $\eta=0.25$, except resting length $a$ is varied. (a) $a\approx 0.449$ is such that there is no tangential stress in steady state; (b) $a\approx 0.899$ is doubled compared to (a), resulting in compressive tangential stress in steady state; (c) $a\approx 0.225$ is halved compared to (a), resulting in tensile tangential stress in steady state.}\label{fig-sin-normalstress}
\end{figure}

Figure~\ref{fig-sin-normalstress} shows snapshots of the discrete model of Figure~\ref{fig-sin} at times $t=0$ and $t=4$ for three different values of spring resting length $a$. Both the tangential stress $\sigma_{\tau\tau}/E$ and the normal stress $\sigma_{nn}/(Eh) = \kappa f(\ell)/(ka)$ are shown. We normalise $\sigma_{nn}$ by Young's modulus $E$ and the height $h$ of the cell layer so that $\sigma_{nn}/(Eh)$ only depends on parameters of the discrete spring model. This normalised normal stress is calculated from Eqs~\eqref{normal-stress}--\eqref{surface-tension} based on the discrete tangential force ${f}(\ell_i)$ exerting on each spring, and the continuous value of curvature along the interface given by
\begin{align}
	\kappa(u) = \frac{y''(u)x'(u) - x''(u)y'(u)}{\Big[\big(x'(u)\big)^2 + \big(y'(u)\big)^2\Big]^{3/2}}
\end{align}
for a parametrisation $\widetilde{\b r}(u)=\big(x(u),y(u)\big)$ of the interface~\citep{sethian1999}.

In Figure~\ref{fig-sin-normalstress}a at time $t=0$, the elongated cell (blue coils) generates compressive (positive) normal stress where the substrate is convex (orange interface region). Similarly, the compressed cell (orange coils) generates compressive normal stress where the substrate is concave. If the spring resting length is increased (Figure~\ref{fig-sin-normalstress}b) or decreased (Figure~\ref{fig-sin-normalstress}c), both the tangential stress and normal stress are modified. At time $t=4$, all cells are relaxed mechanically, so that tangential stress is homogenous along the interface. There is no tangential stress in Figure~\ref{fig-sin-normalstress}a (green coils) since the resting length is chosen such that the springs are at their resting length in mechanical equilibrium. The absence of tangential stress means that there is no normal stress either (green interface). In Figure~\ref{fig-sin-normalstress}b at $t=4$, mechanical equilibrium is reached with compressive tangential stress (orange coils). This compressive tangential stress induces tensile normal stress where the substrate is convex (blue interface region), and compressive normal stress where the substrate is convex (orange interface region). The opposite is found in Figure~\ref{fig-sin-normalstress}c at $t=4$, where there is tensile tangential stress in mechanical equilibrium (blue coils). Tensile tangential stress generates compressive normal stress where the substrate is convex (blue interface region), and tensile normal stress where the substrate is concave (orange--red interface region). In all cases, there is no normal stress wherever curvature is zero (green interface regions).

\section*{Conclusions}
The development of discrete mathematical models of collective cell mechanics in confluent epithelial layers and their continuum limit have provided useful relationships between mechanical cell properties and tissue-scale cell diffusion properties~\citep{murray2009,murray2011,murray2012,fozard2010,murphy2019,murphy2020,lorenzi2020,murphy2021,baker2019,tambyah2020}. These models are particularly important in mechanobiology, where mechanical stress on a cell may influence its differentiation, phenotype, and behaviour~\citep{opas1989,weinans1996,nelson2005,keefer2011,ladoux2017,xi2019,nelson2022}. While the discrete models are easy to formulate and conceptualise, their continuum limit can provide useful analytical insights of the discrete models, such as the emergence of travelling waves~\citep{murphy2021b}. The mechanics and growth of biological tissues is complex due to the fact that they often involve large deformations and differential creation of new material within the tissue~\citep{gamsjaeger2013,goriely2017,ambrosi2019}. Linear elasticity may not apply since biological tissues are often subjected to large stresses, including when cells proliferate and when they experience residual stress in equilibrium. Cell-based mathematical models of tissue growth are helpful in deriving growth laws of the continuum mechanics of biological tissues from first principles.

In the present work, we generalise a simple model of the mechanical interactions of cells arranged along a one-dimensional axis, to curved cells arranged as a chain along an arbitrary parametric curve in two-dimensional space, with the specific aim to understand the influence of curvature for the mechanical relaxation of such cellular tissues. We propose a new derivation of the continuum limit of this discrete model based on expansions in terms of small spring lengths, with well-defined cell densities over finite domains. This new derivation allows us to justify how spring restoring force laws must rescale with the number of springs to obtain consistent dynamics in the limit, and to calculate the order of approximation provided by the linear or nonlinear diffusion equation obtained in the limit. We also provide estimates of the timescale of mechanical relaxation of the tissue based on analysing the discrete models, and justify that numerical simulations of discrete models that assume Lennard--Jones elastic potentials in \citet{murray2012} are linearly unstable.

\cbeg Our findings show that curvature directly influences the mechanical relaxation of cells modelled by a chain of straight springs that bridge across curved regions of the interface. For smooth interfaces, this curvature dependence is weak and decreases with the number of springs per cell $m$ as $\Order\big(m^{-2}\big)$. The rate of convergence of the curved spring model and the straight spring model to the curvature-independent diffusion equation obtained in the continuum limit $m\to\infty$ is $\Order\big(m^{-2}\big)$ in both cases. Cells are soft matter, partly fluid and partly solid, and their fluid nature means that they can deform to take on the curved shape of their substrate, which can be modelled by increasing the number of springs per cell. \cend The mechanical behaviour of a single cell is in itself a complex combination of many dynamic processes that involve actin polymerisation, adhesion complexes in the cell membrane, and hydrostatic pressure of the cytoplasm~\citep{moeendarbary2014,ladoux2017}. While we did not model these intracellular mechanical properties in detail, the straight spring model may represent the mechanical behaviour of actin fibres in the cytoskeleton particularly for transmitting tensile forces. The curved spring model may represent the fact that cytosol-filled cell membranes are curved and can fit substrate curvature, while resisting compressive pressure. To consider these two aspects jointly, more realistic cell body shapes that account for cell thickness would need to be considered. \cbeg However, the fact that the same continuum limit is obtained in both of our discrete models of cellular mechanics suggests that to some extent, this detail may not matter much for tissue scale relaxation dynamics, except when there are low number of cells, and on interfaces with large curvatures that may prevent lateral cell motion, such as cusps in engineered bioscaffolds~\citep{buenzli2020}.\cend

The only quantity that explicitly depends on curvature in the continuum limit is the normal stress of the cells. Tangential forces within the cellular tissue layer generates a surface tension on curved regions of the interface. Biological cells are known to change their behaviour depending on their stress state, so knowing both the tangential and normal components of the stress may be useful for developing mathematical models of cells that account for such behaviours. 

\cbeg In summary, our results suggest important conclusions about modelling cells on curved geometries:
\begin{itemize}
\item Curved and straight springs lead to different dynamics when there is a finite number of springs, which is always unavoidable in computer models;
\item Curved and straight springs converge quadratically to the same dynamics in the continuum limit, if the interface is smooth. This is an important result as it means that strong curvature-dependent behaviours observed experimentally are likely to have other origins than tangential mechanical relaxation;
\item Cells on a curved substrate may sense substrate curvature at length scales much larger than their cell body through normal stress induced by surface tension. This could induce curvature dependences of cell behaviours in experiments.
\end{itemize}
\cend
Our mathematical model focuses on the mechanical relaxation of a cell monolayer in two-dimensions on a static, curved interface, and as such, could be extended in several directions in future works. \cbeg The interaction between dynamic evolutions of the interface with the mechanics of cells is an important consideration for tissue growth and could help explain the strong control of geometry in tissue engineered constructs~\citep{buenzli2020,fratzl2022,karakaya2022}. Cells on moving boundaries experience curvature-controlled crowding and spreading, which can be countered by mechanical relaxation~\citep{alias2017,alias2019,hegartycremer2021}. \cend Normal reaction forces and surface tension may also participate in the evolution of the interface~\citep{ladoux2017,bidan2012a,bidan2012b,bidan2016,fratzl2022}, for example in negative pressure wound therapies that apply negative pressure to the wound area~\citep{huang2014,flegg2020}. Cellular behaviours such as proliferation, death, directed motion along the interface, \cbeg the thickness of the cellular layer, and competition with other tissues are important to include for modelling  more complex, dynamic tissues~\citep{odell1981,cox2018,murphy2019,schamberger2023,cox2024, brown2024}. For example, curvature interacts with the thickness of the cellular layer to generates cells with particular polygonal shapes called scutoids~\citep{gomezgalvez2018, schamberger2023}. \cend

\subsection*{Acknowledgments}
This research was supported by the Australian Research Council (DP190102545, DP230100025). \cbeg We would like to thank the two anonymous reviewers and the Editor for their helpful suggestions\cend.

\bigskip

\begin{appendices}
\section{Continuum limit}\label{appx:continuum-limit}
\begin{figure}
	\centering\includegraphics[scale=0.85]{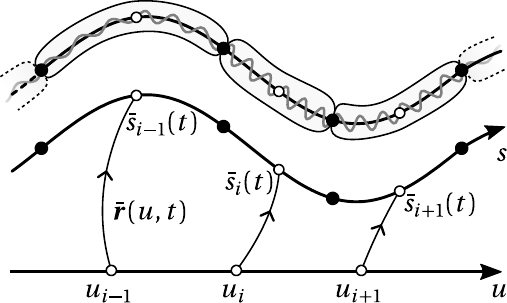}
	\caption{Time-dependent reparametrisation of the interface. The time-dependent parametrisation $\overline{\b r}(u,t)$ of the interface $\b r(s)$ maps constant, evenly spaced coordinates $u_i$ to the time-dependent spring midpoint positions $\b r\big(\overline{s}_i(t)\big)$ (open circles).}
	\label{fig-continuum-limit-reparametrisation}
\end{figure}

To derive the continuum evolution equation~\eqref{evo-spring-density-continuous} from the discrete evolution equation~\eqref{evo-density-midpoint} as $m\to\infty$, we introduce a time-dependent parametrisation $\overline{\b r}(u,t)$ of the interface $\b r(s)$ that tracks the evolving midpoint spring positions for constant parameter values $u$, i.e.,
\begin{align}\label{bar-u}
  \overline{\b r}(u_i, t) = \b r\big(\overline{s}_i(t)\big),
\end{align}
where $u_i = i\Deltaup u$, $i=0,\ldots, M$ are time-independent, evenly spaced coordinates in a finite parameter space $u\in[0,U)$, and $\overline{s}_i(t)$ are the arc length coordinates of the spring midpoints~(Figure~\ref{fig-continuum-limit-reparametrisation}). As $m\to\infty$, $\Deltaup u=U/M\to 0$ and $\overline{s}_{i-1}(t)\sim\overline{s}_i(t)\sim\overline{s}_{i+1}(t)$. This parametrisation of the interface is such that
\begin{align}\label{midpoint-arclength-2}
	\overline{s}_i(t) = \int_0^{u_i}\d u\ g(u,t), \qquad g(u,t) = \norm{\pd{\overline{\b r}}{u}(u,t)}.
\end{align}
Spring density at arc length position $\overline{s}_i(t)$, which corresponds to the coordinate $u_i$, is represented by
\begin{align}
	\rho\big(\overline{s}_i(t),t\big) = \frac{1}{g(u_i,t)\Deltaup u}.\label{rho-g}
\end{align}
Using Eq.~\eqref{midpoint-arclength-2} to express $\overline{s}_{i\pm 1}(t)$ in terms of $u_{i\pm1}=u_i\pm\Deltaup u$ and expanding about $u_i$ as $\Deltaup u\to 0$, we obtain (omitting the time dependence of $\overline{s_i}$ to simplify notation),
\begin{align}\label{midpoint-offsets}
	\overline{s}_{i\pm1} = \overline{s}_i  \pm g(u_i,t)\Deltaup u + \pd{g}{u}(u_i,t)\frac{\Deltaup u^2}{2} \pm \pd{^2g}{u^2}(u_i,t)\frac{\Deltaup u^3}{6} + \Order\big(\Deltaup u^4\big).
\end{align}
With the expressions~\eqref{midpoint-offsets} and $\ell_i^2=1/\rho^2(\overline{s}_i,t) \sim g^2(u_i,t)\Deltaup u^2$, expanding the right hand side of Eq.~\eqref{evo-density-midpoint} about $u_i$ as $\Deltaup u\to 0$ gives
\begin{align}\label{evo-continuum-rhs}
	\td{}{t}\rho\big(\overline{s}_i,t\big) = - \frac{1}{\eta} \pd{^2}{s^2}\widetilde{f}\big(\rho(\overline{s}_i,t)\big) - \frac{1}{\eta} \frac{1}{g^2(u_i,t)}\pd{g}{u}(u_i,t)\pd{}{s}\widetilde{f}\big(\rho(\overline{s}_i,t)\big) + \Order\left(\Deltaup u^2\right), \qquad m\to\infty.
\end{align}
In the left hand side of Eq.~\eqref{evo-continuum-rhs}, we have, using the chain rule,
\begin{align}
  \td{}{t}\rho(\overline{s}_i,t) &= \pd{\rho}{t}(\overline{s}_i,t) + \pd{\rho}{s}(\overline{s}_i,t) \td{\overline{s}_i}{t},
\end{align}
where from Eqs~\eqref{discrete-evo},~\eqref{rho-cont}--\eqref{f-tilde}, and~\eqref{midpoint-offsets},
\begin{align}\label{evo-midpoint}
  \td{\overline{s}_i}{t} = \frac{1}{2\eta}\left[\widetilde{f}\big(\rho(\overline{s}_{i+1},t)\big) - \widetilde{f}\big(\rho(\overline{s}_{i-1},t)\big)\right] = \frac{1}{\eta} \pd{\widetilde{f}}{s}\big(\rho(\overline{s}_i,t)\big) g(u_i,t)\Deltaup u + \Order\left(\Deltaup u^3\right).
\end{align}
Since from Eqs~\eqref{midpoint-arclength-2}--\eqref{rho-g} $g(u_i,t)\Deltaup u = 1/\rho(\overline{s}_i,t)$ and $(\p\rho/\p s)/\rho = -(\p g/\p s)/g = -(\p g/\p u)/g^2$, the left hand side of Eq.~\eqref{evo-continuum-rhs} asymptotically becomes:
\begin{align}\label{evo-continuum-lhs}
  \td{}{t}\rho(\overline{s}_i,t) &= \pd{\rho}{t}(\overline{s}_i,t) - \frac{1}{\eta} \frac{1}{g^2(u_i,t)} \pd{g}{u}(u_i,t)\pd{\widetilde{f}}{s}\big(\rho(\overline{s}_i,t)\big) + \Order\left(\Deltaup u^3\right)
\end{align}
Equating Eqs~\eqref{evo-continuum-lhs} and~\eqref{evo-continuum-rhs} shows that the second term in the right hand sides of both equations cancel, so that
\begin{align}
  \pd{\rho}{t}(\overline{s}_i,t) &= -\frac{1}{\eta}\pd{^2}{s^2}\widetilde{f}\big(\rho(\overline{s}_i,t)\big) \left(1+ \Order\left(\Deltaup u^2\right)\right),\label{evo-spring-density-continuous-appx}
\end{align}
which is the same as Eq.~\eqref{evo-spring-density-continuous}. This development shows that this partial differential equation is a second-order accurate representation of the discrete model equation~\eqref{evo-density-midpoint} as $m\to\infty$, i.e., corrections are $\Order(\Deltaup u^2) = \Order(1/m^2)$.

If $\overline{s}_i(t)$ is not defined as the midpoint between $s_i(t)$ and $s_{i-1}(t)$, corrections of order $\Order(1/m)$ are obtained for the evolution of cell density. Indeed, defining
\begin{align}
	\overline{s}_i(t) = \frac{1+\epsilon}{2} s_i(t) + \frac{1-\epsilon}{2} s_{i-1}(t)  
\end{align}
for $-1\leq \epsilon \leq 1$, such that the midpoint is obtained when $\epsilon=0$, the expansion in Eq.~\eqref{evo-continuum-rhs} remains the same. However, the evolution of $\overline{s}_i(t)$ in Eq.~\eqref{evo-midpoint} now has $\epsilon \Order(\Deltaup u^2)$ corrections. These corrections modify Equation~\eqref{evo-continuum-lhs} into
\begin{align}\label{evo-continuum-lhs2}
  \td{}{t}\rho(\overline{s}_i,t) &= \pd{\rho}{t}(\overline{s}_i,t) - \frac{1}{\eta} \frac{1}{g^2(u_i,t)} \pd{g}{u}(u_i,t)\pd{\widetilde{f}}{s}\big(\rho(\overline{s}_i,t)\big)\notag
  \\&+\frac{\epsilon}{2\eta}\left(\frac{1}{\rho^2(\overline{s}_i,t)}\pd{\rho}{s}(\overline{s}_i,t)\pd{^2}{s^2}\widetilde{f}\big(\rho(\overline{s}_i,t)\big) - \frac{1}{\rho^3(\overline{s}_i,t)}\left(\pd{\rho}{s}(\overline{s}_i,t)\right)^2\pd{}{s}\widetilde{f}\big(\rho(\overline{s}_i,t)\big) \right)\notag
  \\&+ \Order\left(\Deltaup u^3\right)
\end{align}
Equating Eqs~\eqref{evo-continuum-lhs2} and \eqref{evo-continuum-rhs}, substituting $\rho = m q$, and using the definition of $F(q)$ in Eq.~\eqref{restoring-force-contlim}, one obtains
\begin{align}
  \pd{q}{t}(s,t) = &-\frac{1}{\eta}\pd{^2}{s^2}F\big(q(s,t)\big) \notag
  \\&- \frac{\epsilon}{2\eta}\left(\frac{1}{q^2(s,t)}\pd{}{s}q(s,t)\pd{^2}{s^2}F\big(q(s,t)\big) - \frac{1}{q^3(s,t)}\left(\pd{q}{s}(s,t)\right)^2\pd{}{s}F\big(q(s,t)\big)\right)\frac{1}{m}\notag
  \\&+ \Order\left(\frac{1}{m^2}\right).
\end{align}

\end{appendices}

\end{document}